\documentclass[journal,a4paper]{IEEEtran}
\usepackage{cite,multirow}
\usepackage{morefloats}
\usepackage[acronyms,nonumberlist,nopostdot,nomain,nogroupskip]{glossaries}
\usepackage{soul}
\usepackage{amsmath,bbm}
\usepackage{hyperref}
\usepackage{amsfonts,amssymb,graphicx,subfigure,tikz}
\usepackage{etoolbox}
\usepackage{color}
\usepackage{diagbox}
\usepackage{epstopdf}
\usepackage{enumerate}
\usepackage{siunitx}
\usepackage{mathtools}
\usepackage{tikz}
\usepackage[utf8]{inputenc}
\usepackage{pgfplots} 
\usepackage{pgfgantt}
\usepackage{pdflscape}
\usepackage{comment}
\usepackage{bm}
\usepackage{balance}
\usetikzlibrary{arrows, positioning, patterns, patterns.meta}
\pgfplotsset{compat=newest}
\pgfplotsset{plot coordinates/math parser=false}
\usepgfplotslibrary{patchplots, groupplots}
\newlength\fheight
\newlength\fwidth
\pgfplotsset{every axis/.append style={
                    label style={font=\scriptsize},
                    tick label style={font=\scriptsize},
                    legend style={font=\scriptsize}
                    }}

\DeclareMathOperator*{\argmin}{arg\,min}
\newacronym{ca}{CA}{Carrier Aggregation}
\newacronym{3gpp}{3GPP}{3rd Generation Partnership Project}
\newacronym{5g}{5G}{5th generation}
\newacronym{5gc}{5GC}{5G Core}
\newacronym{adc}{ADC}{Analog to Digital Converter}
\newacronym{afbw}{AFBW}{Average Fading Bandwidth}
\newacronym{aimd}{AIMD}{Additive Increase Multiplicative Decrease}
\newacronym{am}{AM}{Acknowledged Mode}
\newacronym{amc}{AMC}{Adaptive Modulation and Coding}
\newacronym{aoa}{AoA}{Angle of Arrival}
\newacronym{aod}{AoD}{Angle of Departure}
\newacronym{ap}{AP}{Access Point}
\newacronym{aqm}{AQM}{Active Queue Management}
\newacronym{awgn}{AGWN}{Additive White Gaussian Noise}
\newacronym{balia}{BALIA}{Balanced Link Adaptation}
\newacronym{bdp}{BDP}{Bandwidth-Delay Product}
\newacronym{ber}{BER}{Bit Error Rate}
\newacronym{bf}{BF}{Beamforming}
\newacronym{bwp}{BWP}{Bandwidth Part}
\newacronym{cad}{CAD}{Computer-Aided Design}
\newacronym{cc}{CC}{Congestion Control}
\newacronym{cdf}{CDF}{Cumulative Distribution Function}
\newacronym{cir}{CIR}{Channel Impulse Response}
\newacronym{cn}{CN}{Core Network}
\newacronym{cp}{CP}{Control Plane}
\newacronym{cqi}{CQI}{Channel Quality Information}
\newacronym{crs}{CRS}{Cell Reference Signal}
\newacronym{csirs}{CSI-RS}{Channel State Information - Reference Signal}
\newacronym{dac}{DAC}{Digital-to-Analog Converter}
\newacronym{dc}{DC}{Dual Connectivity}
\newacronym{dce}{DCE}{Direct Code Execution}
\newacronym{dci}{DCI}{Downlink Control Information}
\newacronym{dl}{DL}{Downlink}
\newacronym{dmr}{DMR}{Deadline Miss Ratio}
\newacronym{dmrs}{DMRS}{DeModulation Reference Signal}
\newacronym{dray}{D-Ray}{Deterministic Ray}
\newacronym{e2e}{E2E}{End-to-End}
\newacronym{ecn}{ECN}{Explicit Congestion Notification}
\newacronym{edf}{EDF}{Earliest Deadline First}
\newacronym{edrx}{eDRX}{Extended Discontinuous Reception}
\newacronym{enb}{eNB}{evolved Node Base}
\newacronym{endc}{EN-DC}{E-UTRAN-\gls{nr} \gls{dc}}
\newacronym{epc}{EPC}{Evolved Packet Core}
\newacronym{es}{ES}{Edge Server}
\newacronym{fdd}{FDD}{Frequency Division Duplexing}
\newacronym{fnbw}{FNBW}{First Null Beam Width}
\newacronym{fdma}{FDMA}{Frequency Division Multiple Access}
\newacronym{fray}{F-Ray}{Flashing Ray}
\newacronym{fs}{FS}{Fast Switching}
\newacronym{ftp}{FTP}{File Transfer Protocol}
\newacronym{gmm}{GMM}{Gaussian Mixture Model}
\newacronym{gnb}{gNB}{Next Generation Node Base}
\newacronym{harq}{HARQ}{Hybrid Automatic Repeat reQuest}
\newacronym{hetnet}{HetNet}{Heterogeneous Network}
\newacronym{hh}{HH}{Hard Handover}
\newacronym{hol}{HOL}{Head-of-Line}
\newacronym{hqf}{HQF}{Highest-quality-first}
\newacronym{ia}{IA}{Initial Access}
\newacronym{iab}{IAB}{Integrated Access and Backhaul}
\newacronym{imt}{IMT}{International Mobile Telecommunication}
\newacronym{inf}{InF}{Indoor Factory}
\newacronym{inf-sh}{InF-SH}{\gls{inf}-Sparse High}
\newacronym{inr}{INR}{Interference to Noise Ratio}
\newacronym{iot}{IoT}{Internet of Things}
\newacronym{ked}{KED}{Knife-Edge Diffraction}
\newacronym{kpi}{KPI}{Key Performance Indicator}
\newacronym{lcf}{LCF}{Level Crossing Frequency}
\newacronym{lcr}{LCR}{Level Crossing Rate}
\newacronym{lo}{LO}{Local Oscillator}
\newacronym{los}{LoS}{Line-of-Sight}
\newacronym{lte}{LTE}{Long Term Evolution}
\newacronym{ltemtp}{LTE-M}{LTE-MTC [Machine Type Communication]}
\newacronym{m2m}{M2M}{Machine to Machine}
\newacronym{mac}{MAC}{Medium Access Control}
\newacronym{mc}{MC}{Monte Carlo}
\newacronym{mcs}{MCS}{Modulation and Coding Scheme}
\newacronym{mec}{MEC}{Mobile Edge Cloud}
\newacronym{mi}{MI}{Mutual Information}
\newacronym{mib}{MIB}{Master Information Block}
\newacronym{mmtc}{mMTC}{massive Machine-Type Communications }
\newacronym{mimo}{MIMO}{Multiple Input Multiple Output}
\newacronym{minlp}{MINLP}{Mixed-Integer Nonlinear Program}
\newacronym{m-mimo}{m-MIMO}{massive-MIMO}
\newacronym{mlr}{MLR}{Maximum-local-rate}
\newacronym{mmwave}{mmWave}{millimeter wave}
\newacronym{moi}{MoI}{Method of Images}
\newacronym{mpc}{MPC}{Multi Path Component}
\newacronym{mptcp}{MPTCP}{Multipath TCP}
\newacronym{mr}{MR}{Maximum Rate}
\newacronym{mrdc}{MR-DC}{Multi \gls{rat} \gls{dc}}
\newacronym{mss}{MSS}{Maximum Segment Size}
\newacronym{mtd}{MTD}{Machine-Type Device}
\newacronym{mtu}{MTU}{Maximum Transmission Unit}
\newacronym{nfv}{NFV}{Network Function Virtualization}
\newacronym{nist}{NIST}{National Institute of Standards and Technology}
\newacronym{nlos}{NLoS}{Non-Line-of-Sight}
\newacronym{nr}{NR}{New Radio}
\newacronym{nrmse}{NRMSE}{Normalized Root Mean Square Error}
\newacronym{nsa}{NSA}{Non Stand Alone}
\newacronym{o2i}{O2I}{Outdoor-to-Indoor}
\newacronym{ofdm}{OFDM}{Orthogonal Frequency Division Multiplexing}
\newacronym{pa}{PA}{Power Amplifier}
\newacronym{prr}{PRR}{Packet Reception Ratio}
\newacronym{pbch}{PBCH}{Physical Broadcast Channel}
\newacronym{pdcch}{PDCCH}{Physical Downlonk Control Channel}
\newacronym{pdcp}{PDCP}{Packet Data Convergence Protocol}
\newacronym{pdsch}{PDSCH}{Physical Downlink Shared Channel}
\newacronym{pdu}{PDU}{Packet Data Unit}
\newacronym{per}{PER}{Packet Error Rate}
\newacronym{pf}{PF}{Proportional Fair}
\newacronym{pgw}{PGW}{Packet Gateway}
\newacronym{phy}{PHY}{Physical}
\newacronym{pl}{PL}{Path Loss}
\newacronym{ppp}{PPP}{Poisson Point Process}
\newacronym{prb}{PRB}{Physical Resource Block}
\newacronym{ps}{PS}{Phase Shifter}
\newacronym{psd}{PSD}{Power Spectral Density}
\newacronym{pss}{PSS}{Primary Synchronization Signal}
\newacronym{pucch}{PUCCH}{Physical Uplink Control Channel}
\newacronym{pusch}{PUSCH}{Physical Uplink Shared Channel}
\newacronym{qam}{QAM}{Quadrature Amplitude Modulation}
\newacronym{qd}{QD}{Quasi Deterministic}
\newacronym{rach}{RACH}{Random Access Channel}
\newacronym{ran}{RAN}{Radio Access Network}
\newacronym[firstplural=Radio Access Technologies (RATs)]{rat}{RAT}{Radio Access Technology}
\newacronym{red}{RED}{Random Early Detection}
\newacronym{redcap}{RedCap}{Reduced Capability}
\newacronym{rf}{RF}{Radio Frequency}
\newacronym{rlc}{RLC}{Radio Link Control}
\newacronym{rlf}{RLF}{Radio Link Failure}
\newacronym{rma}{RMa}{Rural Macro}
\newacronym{rr}{RR}{Round Robin}
\newacronym{rray}{R-Ray}{Random Ray}
\newacronym{rrc}{RRC}{Radio Resource Control}
\newacronym{rrm}{RRM}{Radio Resource Management}
\newacronym{rs}{RS}{Remote Server}
\newacronym{rsrp}{RSRP}{Reference Signal Received Power}
\newacronym{rsrq}{RSRQ}{Reference Signal Received Quality}
\newacronym{rss}{RSS}{Received Signal Strength}
\newacronym{rssi}{RSSI}{Received Signal Strength Indicator}
\newacronym{rt}{RT}{Ray Tracer}
\newacronym{rtt}{RTT}{Round Trip Time}
\newacronym{rw}{RW}{Receive Window}
\newacronym{rx}{RX}{Receiver}
\newacronym{sa}{SA}{standalone}
\newacronym{sack}{SACK}{Selective Acknowledgment}
\newacronym{sap}{SAP}{Service Access Point}
\newacronym{sch}{SCH}{Secondary Cell Handover}
\newacronym{scm}{SCM}{Spatial Channel Model}
\newacronym{scoot}{SCOOT}{Split Cycle Offset Optimization Technique}
\newacronym{sdma}{SDMA}{Spatial Division Multiple Access}
\newacronym{sf}{SF}{Shadow Fading}
\newacronym{si}{SI}{Study Item}
\newacronym{sib}{SIB}{Secondary Information Block}
\newacronym{sinr}{SINR}{Signal-to-Interference-plus-Noise Ratio}
\newacronym{sir}{SIR}{Signal-to-Interference Ratio}
\newacronym{sm}{SM}{Saturation Mode}
\newacronym{sma}{SmA}{Smart Agriculture}
\newacronym{snr}{SNR}{Signal-to-Noise Ratio}
\newacronym{son}{SON}{Self-Organizing Network}
\newacronym{srs}{SRS}{Sounding Reference Signal}
\newacronym{ss}{SS}{Synchronization Signal}
\newacronym{ssb}{SSB}{Synchronization Signal Block}
\newacronym{sss}{SSS}{Secondary Synchronization Signal}
\newacronym{sta}{STA}{Station}
\newacronym{tb}{TB}{Transport Block}
\newacronym{tcp}{TCP}{Transmission Control Protocol}
\newacronym{udp}{UDP}{User Datagram Protocol}
\newacronym{tdd}{TDD}{Time Division Duplexing}
\newacronym{tdma}{TDMA}{Time Division Multiple Access}
\newacronym{tfl}{TfL}{Transport for London}
\newacronym{tgad}{TGad}{Task Group ad}
\newacronym{tgay}{TGay}{Task Group ay}
\newacronym{tm}{TM}{Transparent Mode}
\newacronym{trp}{TRP}{Transmitter Receiver Pair}
\newacronym{tti}{TTI}{Transmission Time Interval}
\newacronym{ttt}{TTT}{Time-to-Trigger}
\newacronym{tx}{TX}{Transmitter}
\newacronym{ue}{UE}{User Equipment}
\newacronym{ul}{UL}{Uplink}
\newacronym{um}{UM}{Unacknowledged Mode}
\newacronym{uma}{UMa}{Urban Macro}
\newacronym{uml}{UML}{Unified Modeling Language}
\newacronym{utc}{UTC}{Urban Traffic Control}
\newacronym{vm}{VM}{Virtual Machine}
\newacronym{wbf}{WBF}{Wired Bias Function}
\newacronym{wf}{WF}{Wired-first}
\newacronym{wifi}{Wi-Fi}{Wireless Fidelity}
\newacronym{wigig}{WiGig}{Wireless Gigabit}
\newacronym{wlan}{WLAN}{Wireless Local Area Network}
\newacronym{xpr}{XPR}{Cross Polarization Ratio}
\newacronym{fr2}{FR2}{Frequency Range 2}
\newacronym{fr1}{FR1}{Frequency Range 1}
\newacronym{nbiot}{NB-IoT}{Narrowband-IoT}
\newacronym{cps}{CPS}{Cyber-Physical production System}
\newacronym{iiot}{IIoT}{Industrial Internet of Things}
\newacronym{agv}{AGV}{Autonomous Ground Vehicle}
\newacronym{uav}{UAV}{Unmanned Autonomous Vehicle}
\newacronym{amr}{AMR}{Autonomous Mobile Robots}
\newacronym{wsn}{WSN}{Wireless Sensor Network}
\newacronym{embb}{eMBB}{enhanced Mobile Broadband}
\newacronym{urllc}{URLLC}{Ultra-Reliable Low-Latency Communications}
\newacronym{lpwan}{LPWAN}{Low-Power Wide Area Network}
\newacronym{lora}{LoRa}{Long Range}
\newacronym{qos}{QoS}{Quality of Service}
\newacronym{ec}{EC}{energy consumption}
\newacronym{bs}{BS}{Base Station}
\newacronym{ula}{ULA}{Uniform Linear Array}
\usepackage[linesnumbered, ruled]{algorithm2e}
\graphicspath{{Figures/}}

\newcommand{\Summary}{\noindent\paragraph*{Summary}}

\begin{document}             % End of preamble and beginning of text.
	%\AmS-\LaTeX
\title{Optimizing Energy Efficiency of 5G RedCap Beam Management for Smart Agriculture Applications}

\author{Manishika~Rawat, Matteo~Pagin,~\IEEEmembership{Student Member,~IEEE}, Marco Giordani,~\IEEEmembership{Member,~IEEE}, \\ Louis-Adrien Dufrene, Quentin Lampin, Michele~Zorzi,~\IEEEmembership{Fellow,~IEEE}

\thanks{Manishika Rawat was with the Department of Information Engineering (DEI) of the University of Padova, Italy. She is now with the Department of Electronics and Communication Engineering, IIT Roorkee, India. Email: manishika.rawat8@gmail.com 

Matteo Pagin, Marco Giordani, and Michele Zorzi are with the Department of Information Engineering (DEI) of the University of Padova, Italy. Email: \{paginmatte,giordani,zorzi\}@dei.unipd.it.

Louis-Adrien Dufrene and Quentin Lampin are with Orange Labs, France. Email: \{louisadrien.dufrene,quentin.lampin\}@orange.com.

A preliminary version of this work was presented at the IEEE Global Communications Conference 2023~\cite{10437644}.

This work was partially supported by the European Union under the Italian
National Recovery and Resilience Plan (NRRP) of NextGenerationEU,
partnership on “Telecommunications of the Future” (PE0000001 - program
“RESTART”).}}

\maketitle
\begin{abstract}
Beam management in 5G NR involves the transmission and reception of control signals such as \glspl{ssb}, crucial for tasks like initial access and/or channel estimation. However, this procedure consumes energy, which is particularly challenging to handle for battery-constrained nodes such as RedCap devices. Specifically, in this work we study a mid-market \gls{iot} \gls{sma} deployment where an \gls{uav} acts as a base station ``from the sky'' (UAV-gNB) to monitor and control ground \glspl{ue} in the field. 
%Specifically, we propose guidelines to optimize the energy efficiency of the beam management procedure at the \gls{uav}-\gls{gnb} while ensuring some \gls{qos} requirements at the mobile ground \glsls{ue}. 
Then, we formalize a multi-variate optimization problem to determine the optimal beam management design for RedCap \gls{sma} devices in order to reduce the energy consumption at the UAV-gNB. %while ensuring some \gls{qos} requirements at the mobile ground \glsls{ue}. 
Specifically, we jointly optimize the transmission power and the beamwidth at the UAV-gNB.  %in \gls{sma} scenario. 
Based on the analysis, we derive the so-called ``regions of feasibility,'' i.e., the upper limit(s) of the beam management parameters for which RedCap \gls{qos} and energy constraints are met. We study the impact of factors like the total transmission power at the gNB, the \gls{snr} threshold for successful packet decoding, the number of \glspl{ue} in the region, and the misdetection probability. Simulation results demonstrate that there exists an optimal configuration for beam management to promote energy efficiency, which depends on the speed of the \glspl{ue}, the beamwidth, and other network parameters.

%In this work, we explored the 5G NR beam management design for RedCap devices in an SmA scenario. We consider a mobile UE which may lose alignment with the associated beam during beam management, potentially resulting in misdetection. Therefore, we formalized an optimization problem to minimize the energy consumption during beam management, while ensuring that some desired \gls{qos} requirements such as the SNR threshold at UE and misdetection probability, are met. We obtain an optimal number of antenna elements for a ULA antenna design and the optimal transmission power at the gNB node large enough to meet the constraints. The impact of the factors like the total transmission power at the gNB node, the SNR threshold, the number of UEs in the served region and misdetection probability is studied. Through simulations, we identified the feasibility regions where the problem can be solved, and proposed the optimal values of the beam management parameters for RedCap devices, such as the optimal SSB size and periodicity, to maintain a minimum energy consumption while optimizing latency and overhead. 
\end{abstract}
\begin{IEEEkeywords}
	5G NR, 3GPP, beam management, energy consumption, optimization, smart agriculture.
\end{IEEEkeywords}

\begin{picture}(0,0)(0,-440)
	\put(0,0){
	\put(0,0){\qquad \qquad \quad This paper has been submitted to IEEE for publication. Copyright may change without notice.}}
\end{picture}

\glsresetall
\section{Introduction}
%Basically a rephrased intro of the GLOBECOM paper

As the world becomes increasingly interconnected, the \gls{3gpp} \gls{nr} standard~\cite{38300} for \gls{5g} networks represents a significant advancement in telecommunications, offering unprecedented high speed, reliability, and connectivity~\cite{itu-r-2083}.
Specifically, 5G \gls{nr} rests on three main service pillars, catering to specific requirements and applications~\cite{22261}: (i) \gls{embb}, providing faster data speeds (up to 20 Gbps in ideal conditions) and greater capacity (up to 10 Mbit/s/m$^2$) compared to previous generations, e.g., for high-definition video streaming or immersive AR/VR experience; (ii) \gls{urllc} supporting mission-critical applications (i.e., with around 1 ms round-trip delay) like autonomous vehicles or remote surgery; and (iii) \gls{mmtc}, supporting interconnected networks of low-cost low-energy-consumption \gls{iot} sensors (with up to $10^6$ connections per km$^2$) that communicate and exchange data.

In particular, \gls{iot} applications span various sectors, from smart homes and cities to industrial automation (e.g., transportation and logistics) and healthcare (e.g., in smart hospitals and/or to facilitate automatic data collection and sensing)~\cite{atzori2010internet,zanella2014internet}.
Notably, the main requirements of \gls{iot} services are long transmission range (in the order of a few kilometers) and low energy consumption to support prolonged battery lifetime for sensors (up to 10 years).
Along these lines, standardization bodies and industry players have contributed to develop various \gls{lpwan} technologies, such as \gls{lora}~\cite{magrin2017performance}, \gls{nbiot}~\cite{matz2020systematic}, and SigFox~\cite{ribeiro2018outdoor}, to provide a good balance between range and energy consumption~\cite{ayoub2018internet} for low-cost low-complexity \gls{iot} use cases.
However, the data rate of \gls{lpwan} systems is generally limited to a few hundreds of Kbps, which may not be compatible with the requirements of future IoT use cases, for example \gls{inf} and/or \gls{sma} applications. \gls{sma}, in particular, involves the use of IoT devices such as sensors, drones, and robots, to generate and distribute data to monitor, manage, and automate various agricultural processes (e.g., crop data collection and/or livestock monitoring)~\cite{8066090}. In this case, data rate requirements can be up to 100 Mbps~\cite{22804} (e.g., when data is sent to optimize/assist farmers' decisions), and the latency can be in the order of a few ms. 

\begin{figure*}
	\centering  
 \setlength\belowcaptionskip{-0.8cm}
	\includegraphics[width=.9\textwidth]{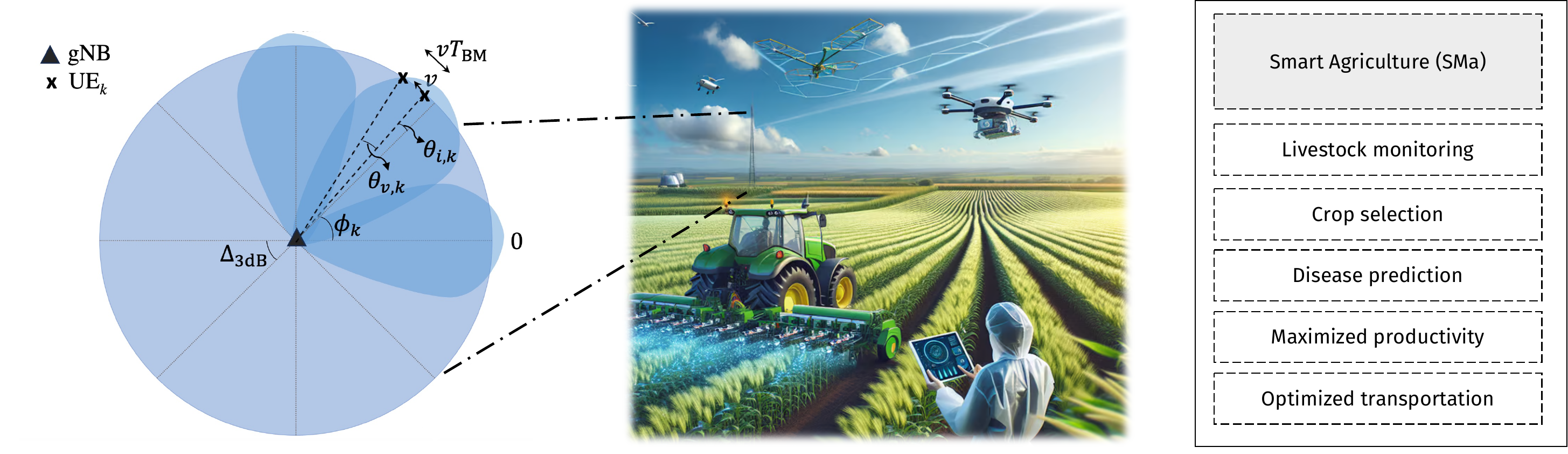}
	\caption{UE mobility model (left) and SMa scenario (right). During beam management, UE$_k$ accumulates an angular offset ${\theta}_k$ due to initial misalignment ($\theta_{i,k}$) and mobility ($\theta_{v,k}$).}
	\label{vel_model}
\end{figure*}

In this context, the \gls{3gpp} is actively promoting a new standard called \gls{redcap}~\cite{38875} to accommodate high-end \gls{iot} devices~\cite{varsier20215g}. 
Specifically, RedCap is positioned to satisfy higher data rate and reliability, and lower latency than current \gls{lpwan} technologies, while promoting lower cost and complexity, longer battery life, and wider coverage than full-blown 5G NR solutions~\cite{38300}. 

As in 5G NR, RedCap devices may operate in both \acrlong{fr1} (between 410 and 7125 MHz) and \acrlong{fr2} (between 24.25 and 52.6 GHz), i.e., in the lower part of the  \gls{mmwave} spectrum, to improve network performance~\cite{pagin2023nrlight}.
However, \glspl{mmwave} come with strict limitations in terms of propagation (mainly severe path loss and absorption), which requires the endpoints to communicate through highly-directional transmissions via \gls{mimo} antenna arrays and beamforming. As such, the transmitter and the receiver need to establish and maintain precise beam alignment for communication, which may increase the control overhead. Along these lines, most prior works study algorithms that periodically align the transmitting and receiving beams to optimize the link quality~\cite{chiu2019active, 9605254, scalabrin2018beam, 8445969}. For instance, Hussain \emph{et al.}~\cite{9605254} proposed a dual-timescale learning-based algorithm to optimize the spectral efficiency by predicting beam dynamics and thus reducing the control overhead. Similarly, Scalabrin \emph{et al.}~\cite{scalabrin2018beam} proposed an adaptive scheme to reduce the beam training overhead by controlling the beamwidth.

% \begin{figure*}
% 	\centering  
%  \setlength\belowcaptionskip{-0.8cm}
% 	\includegraphics[width=.9\textwidth]{Figures/SMa.pdf}
% 	\caption{UE mobility model. During beam management, UE$_k$ accumulates an angular offset ${\theta}_k$ due to both initial misalignment ($\theta_{i,k}$) and mobility ($\theta_{v,k}$).}
% 	\label{vel_model}
% \end{figure*}

In 5G \gls{nr}, beam alignment is determined and preserved by a control procedure referred to as beam management~\cite{giordani2019tutorial}. 
In this process, the \gls{gnb} continuously sends control signals at predefined intervals and directions in the form of \glspl{ssb}, possibly grouped into bursts. Upon receiving and detecting the \glspl{ssb}, end users can identify the strongest beam to connect to, and the resulting direction of transmission towards the \gls{gnb}. 
However, beam management consumes significant energy for sending and receiving control signals, especially when using narrow beams and short \gls{ssb} periodicity~\cite{giordani2017improved}.
While this might not be a critical concern for 5G \gls{nr} systems, it may be challenging for low-complexity battery-powered RedCap devices, thus potentially degrading network~performance.
To date, only a few works explored the complexity and power consumption of 5G NR beam management for IoT (though not specifically for RedCap). %e.g., ~\cite{8590817, 9773088, mukherjee2020energy, hussain2018energy}. 
For example, Zhao \emph{et al.}~\cite{8590817} designed an efficient beam training scheme for IoT devices, although the evaluation is only in terms of the system capacity. Zeulin \emph{et al.}~\cite{9773088} used a digital twin of an industrial IoT environment to predict beam dynamics and thus simplify the beam scanning procedure. Mukherjee \emph{et al.}~\cite{mukherjee2020energy} studied the energy consumption of beam management in the context of spectrum sharing.

Recent efforts in the scientific community have explored several ways to simplify the 5G NR standard to optimize power consumption for RedCap devices~\cite{pagin2023nrlight}. These include, for example: (i) a simpler \gls{mimo} design at the user terminals, with much smaller form factor and fewer layers; (ii) the use of low-cost hardware components; (iii) the use of a lower bandwidth compared to 5G NR, ranging from 50 to 100 MHz; (iv) relaxation of the maximum modulation order, from 6 to 4; and (v) the introduction of power saving functionalities, such as \gls{edrx} or wake-up signals. %\footnote{The \gls{3gpp} has initiated Study and Work Items, including TR 38.869~\cite{38869}, to investigate low-power wake-up signals and receivers specifically designed for RedCap devices.}
However, to the best of our knowledge, there has been little research specifically focusing on how to optimize control network procedures for RedCap, including beam management. % which demands for additional effort in this~sense.

In our previous work~\cite{10437644}, we addressed this issue in an \gls{inf} scenario. Specifically, we described an algorithm to minimize the energy consumption during beam management by optimizing the number of antennas at the gNB, with constraints related to the maximum transmission power at the gNB and the probability of misdetection for the user terminals. However, we did not optimize the transmission power at the gNB. In fact, while for an \gls{inf} scenario the gNB is generally connected to the electric grid and may not be power limited, there are some other scenarios where instead power is a critical parameter. Consider, for example, an \gls{sma} use case, as represented in Fig.~\ref{vel_model}, in which a \gls{uav} acts as a gNB ``from the sky'' to collect data from sensors in the field, like in vineyards (e.g., one sensor per vine) or in plantations (e.g., one sensor per fruit tree)\footnote{Although most RedCap energy and/or complexity reduction techniques concern UE-side procedures, the RedCap framework is typically considered as an additional 5G use case targeting ``devices" rather than UEs only~\cite{ericcson2023}. Therefore, simplifications at the gNB are also included here.}.

%This paradigm is being promoted by various international organizations, e.g., the agricultural European Innovation Partnership (EIP-AGRI). 
Now, \glspl{uav} use the battery for propulsion and hovering~\cite{traspadini2023energy}, %and the flight time is generally limited to a few tens of minutes~\cite{traspadini2023energy}. 
%In this scenario,
so it becomes particularly important to minimize the energy consumption under several aspects, including at the control plane for network operations such as beam~management.

To bridge these research gaps, in our paper we extend our previous work in~\cite{10437644} by optimizing the energy consumption of beam management for RedCap devices in an \gls{sma} scenario. Our contributions can be summarized as follows:
\begin{itemize}
    \item We formalize a multi-variate optimization problem to minimize the energy consumption of RedCap beam management. We consider an aerial gNB onboard a \gls{uav} monitoring soil and/or supporting harvesting or pest monitoring of live stocks. Then, we jointly optimize the number of antennas (therefore the beamwidth) and the transmission power at the \gls{uav}-\gls{gnb}. We set constraints in terms of the maximum transmission power that can be used at the \gls{uav}-\gls{gnb} to sustain the flight, and the misdetection probability for the ground terminals/sensors.
    \item As our optimization problem is non-convex, we propose to solve it via a \gls{mc} algorithm. In particular, we find the lower bound of the expression of the energy consumption of beam management, and show that its derivative is non-negative, so that our optimization problem chooses the minimum values of the optimization variables that meet the problem's constraints.
    \item We run a vast simulation campaign to dimension beam management for RedCap devices in an \gls{sma} scenario. Numerical results are given as a function of many parameters, including the speed of the ground terminals, the number of \glspl{ssb} per burst and the burst periodicity, the maximum transmission power at the UAV-gNB, and the number of ground terminals. We demonstrate that there exists an optimal configuration for beam management that can minimize the energy consumption at the UAV-gNB.
    \item We derive the so-called ``regions of feasibility,'' i.e., the upper limits of the beam management parameters, such as the number of \glspl{ssb} per burst and the burst periodicity, where \gls{qos} constraints are met, e.g., in terms of misdetection probability. Based on that, we provide guidelines on the optimal beam management design for RedCap devices in an \gls{sma} scenario.
\end{itemize}

The rest of the paper is organized as follows. In Sec.~\ref{sec2} we present our system model (deployment, energy, mobility, and beam management). In Sec.~\ref{sec3} we describe our optimization problem, the impact of the number of antenna elements at the \gls{gnb} and the transmission power on the \gls{qos} constraints, and provide some analytical results. In Sec.~\ref{results} we present the simulation results and provide design guidelines towards the optimal set of parameters for beam management. Finally, conclusions are given in Sec.~\ref{conclude}.

\section{System Model}\label{sec2}
In this section we present our deployment (Sec.~\ref{sub:deployment}), beam management (Sec.~\ref{sub:beam-management}), energy consumption (Sec.~\ref{sub:ec}), and beam dynamics (Sec.~\ref{sub:mob}) models.

\subsection{Deployment Model}
\label{sub:deployment}

We consider a \gls{3gpp} \gls{rma} scenario~\cite{3GPPrel16} with a radius $R$, served by a \gls{uav} acting as a \gls{gnb} and deployed at height $h_{\rm gNB}$. The $K$ RedCap \glspl{ue} of height $h_{\rm UE}$ are distributed uniformly in the serving region. The location of UE$_k$, for $k\in \{1, 2, \dots, K\}$, is given by the polar coordinates $(d_k, \phi_k)$, where $d_k$ is the 2D distance between UE$_k$ and the gNB, and $\phi_k$ is the angle of UE$_k$ with respect to the positive x-axis, measured counterclockwise. 
The \glspl{ue} are assumed to be moving on a circle at constant velocity $v$ in a counterclockwise direction.\footnote{
Notice that the mobility model is defined only in terms of the tangential velocity $v$. However, in this paper we focus on the probability that a UE moves outside of the coverage of a certain beam, which indeed only depends on $\phi_k$ and so on the tangential velocity. In fact, the normal component of the velocity has an impact only on $d_k$, and is thus neglected in this work.} %MP: Added this since we are likely going to get reviewers' comments about this assumption otherwise (already happened for the conf. paper)

If $P_T$ is the transmission power at the gNB, then the average \gls{snr} at UE$_k$ can be expressed as $P_T\gamma_k$. In turn, $\gamma_k$ can be defined as~\cite{rawat2023optimal}
\begin{equation}
\gamma_{k}(d_{\rm3D})=
\frac{\mathcal{H}_{\rm L}P_r(d_{\rm 3D})+\mathcal{H}_{\rm N}(1-P_r(d_{\rm 3D}))}{N_0\cdot B /G_{\text{gNB},k}G_{\rm UE}},
\label{gamma_k}
\end{equation}
where $d_{\rm 3D}=\sqrt{(h_{\rm gNB}-h_{\rm UE})^2+d_k^2}$ is the distance between the gNB and UE$_k$, $N_0$ is the noise \gls{psd} and $B$ is the channel bandwidth. 
$P_r(d_{\rm 3D})$ is the \gls{los} probability, as described in~\cite{3GPPrel16}, and $\mathcal{H}_{j}$ represents the joint effect of path loss, shadowing and fading and is thus defined as

\begin{equation}
\mathcal{H}_{j}=|\mathbbm{h}_{j}^{k}|^2 / \text{PL}_{j}^{k},\: j\in\{\text{L,N}\},
\end{equation}
where $\mathbbm{h}_{j}^{k}$ and PL$_{j}^k$ are the channel fading gain and path loss for the \gls{los} (L) and \gls{nlos} (N) links, respectively.
Finally, in Eq.~\eqref{gamma_k}, $G_{\text{gNB},k}$ ($G_{\text{UE}}$) represents the beamforming gain at the gNB  (UE) towards (of) the $k$-th user.
We assume analog beamforming (a realistic assumption for RedCap devices to minimize energy consumption~\cite{abbas2017}) at the gNB, which can thus probe only one direction at a time. 
Specifically, the gNB is equipped with a uniform linear  antenna array with $N_{\rm gNB}$ elements, so the beamforming gain is expressed as~\cite{balanis}
	\begin{align}
		G_{\text{gNB},k}=\left|\sin\left(\dfrac{\pi N_{\rm gNB}}{2}\sin{\theta}_k\right)/\sin\left(\dfrac{\pi}{2}\sin{\theta}_k\right)\right|,
  \label{gain_gNB}
	\end{align}
where $\theta_{k}$ is the angular offset with respect to UE$_k$, as described in Sec.~\ref{sub:mob}. Moreover, we assume that \glspl{ue} are equipped with a single isotropic antenna to promote energy efficiency, i.e., $G_{\text{UE}} \equiv 1$.

\subsection{Beam Management Model}
\label{sub:beam-management}
According to the 5G NR specifications~\cite{38300}, beam management relies on a directional version of the 4G LTE synchronization signal called \gls{ssb}.
Specifically, each SSB consists of 4 OFDM symbols in time and $240$ subcarriers in frequency, where the subcarrier spacing depends on the 5G NR numerology \cite{giordani2019tutorial}. 
Each SSB is mapped into a certain angular direction so that directional measurements can be made based on the quality of the received signal, e.g., in terms of the \gls{snr}.
To reduce the overhead, SSBs are grouped into SS bursts, which consist of $N_{\rm SS}\in \{8, 16, 32, 64\}$ SSBs, contiguous in time. The periodicity between consecutive SS bursts is $T_{\rm SS}\in \{5, 10, 20, 40, 80, 160\}$ ms. 

\subsection{Energy Consumption Model}
\label{sub:ec}

In 5G NR beam management, the gNB transmits the SSBs by sequentially sweeping different angular directions to cover the whole beam space (or cell sector).
The transmission of these control signals entails an energy consumption at the gNB which can be expressed as
\begin{equation}
	E_C=S_{ D} P_{\rm gNB}T_{\rm SSB},
	\label{EC1}
\end{equation}
where $S_{ D}$ is the number of SSBs required to completely sweep the beam space (which is a function of the beamwidth at the gNB), $P_{\rm gNB}$ is the power consumed for transmitting each SSB, and $T_{\rm SSB}$ is the time required to send each SSB.

From \cite[Eq. (3)]{giordani2019tutorial}, $S_{ D}$ relative to the horizontal plane (with azimuth ranging from $0$ to $2\pi$) can be expressed~as 
 %the number of SSBs required to completely sweep the beam space on the horizontal plane, with azimuth ranging from $0$ to $2\pi$, can be expressed~as 
\begin{align}
	S_D=\left\lceil{2\pi}/{\Delta_{3 \rm dB}}\right\rceil \approx \lceil\pi N_{\rm gNB}\rceil,
	\label{eq:SD}
\end{align}
where $\Delta_{3 \rm dB}\approx 2/N_{\rm gNB}$ is the 3-dB beamwidth~\cite{balanis}.
%Thus, if $N=4$, then approximately $13$ beams will be required to completely sweep all the directions in the InF beam space. 
Since each SSB consists of 4 OFDM symbols, the time (in $\mu$s) required to send one SSB can be expressed as \cite[Eq. (2)]{giordani2019tutorial}
\begin{align}
	T_{\rm SSB}=4T_{\rm symb}=4\left({71.45}/{2^n}\right),
\end{align}
where $n$ represents the 5G NR numerology index.

The transmitter RF front-end for analog beamforming consists of a pair of \glspl{dac} (one for each I/Q channel), one RF chain, a \gls{lo}, a \gls{pa}, and $N_{\rm gNB}$ \glspl{ps}. According to \cite{energyefficiency2018}, the total power consumption at the transmitter is thus given by
\begin{equation}
	P_{\rm gNB}=N_{\rm gNB}P_{\rm PS}+P_{\rm PA}+P_{\rm RF}+P_{\rm LO}+2P_{\rm DAC},
	\label{power_gNB}
\end{equation}
where $P_{\rm RF}=2P_{\rm M}+2P_{\rm LPF}$ is the power consumption of the RF chain, and $P_{\rm PA}$ stands for the power consumed by the power amplifier. In turn, $P_{\rm PA}$ can be expressed as $P_{\rm PA}=P_T/\eta$, where $\eta$ is the power-added efficiency~\cite{poweramp2017}. 
$P_{\rm DAC}$  
%(b_{\rm DAC}, F_s )$ 
denotes the power consumption of a \gls{dac}, and can be computed as~\cite[Eq. (13)]{energyefficiency2018}
\begin{align}
P_{\rm DAC}=1.5\times 10^{-5}2^{b_{\rm DAC}}+9\times 10^{-12}b_{\rm DAC}F_s,
\end{align}
where $F_s$ is the sampling frequency and $b_{\rm DAC}$ is the number of resolution bits. 
A thorough description of the power components appearing in Eq.~\eqref{power_gNB}, and the corresponding numerical values used in this work, is provided in Table~\ref{power_table}.
\begin{table}[t!]
\centering
\footnotesize
\renewcommand{\arraystretch}{1.1}
\caption{Power consumption parameters.}
\label{power_table}
	\begin{tabular}{ l|l|l }
		\hline
		{Parameter} & {Description} & {Value}\\
		\hline
 		$P_{\rm PS}$&  Phase shifter & $21.6$ mW\\
 		$P_M$ & Mixer & $0.3$ mW\\
 		$P_{\rm LO}$    &Local oscillator &  $22.5$ mW\\
 		$P_{\rm LPF}$&  Low pass filter & $14$ mW\\
 		$P_{\rm BB}$& Baseband amplifier & $5$ mW\\
 		$P_{\rm DAC}$& DAC & \cite[Eq. (13)]{energyefficiency2018}\\
		\hline
	\end{tabular}
\end{table}

% \begin{table}[t!]
% \centering
% \footnotesize
% \renewcommand{\arraystretch}{1.1}
% \caption{Power consumption parameters.}
% \label{power_table}
% 	\begin{tabular}{ l|l }
% 		\hline
% 		{Parameter} & {Description}\\
% 		\hline
%  		$\Delta_{3\rm dB}$ & 3-dB beamwidth \\
%  		$\phi_k$ & UE angle \\
%  		$\theta_{v,k}$ & Movement angular offset \\
%  		$\theta_{i,k}$ & Initial angular offset \\
%  		$v$ & Speed of the UE\\
%  		$T_{\rm BM}$ & Beam management time\\
% 		\hline
% 	\end{tabular}
% \end{table}

\subsection{Beam Dynamics Model}
\label{sub:mob}
At first, UE$_k$, $k\in \{1, 2, \dots, K\}$, establishes an initial connection with the gNB using beam
\begin{equation}
m_{i,k} = \argmin\limits_m (\phi_k-\phi_{br, m}), \; m \in\left\{ 1, 2, \dots, S_D \right\},
\end{equation}
where $\phi_{br, m} \equiv (m-1) \Delta_{3\rm dB}$ is the antenna boresight direction of the $m$-th beam.
Due to the finite nature of the pre-defined codebook of directions available at the gNB, UE$_k$ thus comes with a non-zero initial angular offset 
\begin{align}
\theta_{i,k}= \phi_k-\phi_{br, m_{i,k}}
\label{theta_ik}
\end{align} 
with respect to the boresight direction of the $m_{i,k}$-th beam, %, given by $\phi_{br,m_{i,k}}=(m_{i,k}-1) \Delta_{3\rm dB}$, 
as represented in Fig.~\ref{vel_model}.
%\hl{MR: Do you think $\phi_{br,m_{i,k}}$ should be defined after eq. (9), instead of after eq. (10)?} 
%This computation determines the beam with which a given UE$_k$ is associated.

Moreover, we assume that during beam management UE$_k$ moves with a tangential speed $v$.
%in a counterclockwise direction at a constant velocity $v$. MP I'd avoid repeating this assumption here. It's already mentioned in the system model
%During this tim, UE$_k$ may lose beam alignment and the corresponding beamforming gain, move out of the associated beam, and get disconnected if the resulting SNR is lower than a pre-defined threshold~\cite{giordani2018coverage}. 
Therefore, during the time in between beam updates, UE$_k$ progressively loses beam alignment, and so the resulting beamforming gain starts to deteriorate.
% and thus experiences a degraded beamforming gain. 
Eventually, UE$_k$ may also end up in the coverage region of a different beam, and possibly get disconnected whenever the resulting SNR is below a given threshold~\cite{giordani2018coverage}.
To account for this misalignment, we define $\theta_{v,k}$, also represented in Fig.~\ref{vel_model}, as the angular offset accumulated due to the movement of  UE$_k$ in between consecutive beam updates, i.e.,
\begin{equation}
	\theta_{v,k}={v T_{\rm BM}}/{d_k}.
	\label{theta_vk}
\end{equation}
In Eq.~\eqref{theta_vk}, $T_{\rm BM}$ is the beam management time, i.e., the time it takes to send SSBs across all the $S_D$ angular directions, which in turn can be expressed as~\cite[Eq.~(4)]{giordani2019tutorial}
\begin{align}
	T_{\rm BM}=T_{\rm SS}\left(\left\lceil{S_D}/{N_{\rm SS}}\right\rceil-1\right)+T_{\ell},
	\label{T_BM}
\end{align}
where $T_{\ell}$ is the time required to send the remaining SSBs in the last burst, and is given in \cite[Eq.~(6)]{giordani2019tutorial}. 
\begin{comment}
\begin{align}
	T_{\ell}=
	\begin{cases}
		\dfrac{N_{\rm SS,\ell}}{2}T_{\rm slot}-2T_{\rm symb} & \hspace{-0.2cm} \text{if } (N_{\rm SS,\ell})\hspace{-0.3cm}\mod2=0;\\[10pt]
		\left\lfloor\dfrac{N_{\rm SS,\ell}}{2}\right\rfloor T_{\rm slot}+6T_{\rm symb} & \hspace{-0.1cm}\text{otherwise}.
	\end{cases}
\end{align}

Specifically, $T_{\rm slot}$ ($T_{\rm symb}$) is the slot (symbol) duration, and $N_{\rm SS, \ell}$ is the number of remaining SSBs in the last burst, i.e., 
\begin{equation}
N_{\rm SS, \ell}=S_D-N_{\rm SS}\left(\left\lceil\dfrac{S_D}{N_{\rm SS}}\right\rceil-1\right).	
\end{equation}
\end{comment}

Therefore, the overall angular offset for UE$_k$ during beam management due to both the initial offset and the offset accumulated due to mobility can be expressed as 
\begin{equation}
	\theta_k=|\theta_{i,k}+\theta_{v,k}|.
	\label{theta_k}
\end{equation}
%\hl{When $S_D<N_{\rm SS}$, it indicates that a single burst will accommodate the required number of beams for the beam management. This implies that $T_{\rm BM}=T_l$ which is a very small value. In such a case, $\theta_{k}$ is mostly dominated by $\theta_{i,k}$}.

\section{Energy Consumption Optimization Problem}\label{sec3}
In Sec.~\ref{sub:baseline_problem} we define an optimization problem P1 to minimize the energy consumption of beam management at the UAV-gNB by tuning both the transmission power and the beamwidth at the gNB, while ensuring a sufficient link quality. Moreover, in Sec.~\ref{sub:non-zero} we generalize the problem to an arbitrary misdetection probability (P2). Finally, in Sec.~\ref{sec:opt_alg} we present our optimization algorithm to solve P1 and P2. %Finally, we provide a closed-form expression of the angular offset in Sec.~\ref{antenna_Gain}, and define the feasibility limits of the problems in Sec.~\ref{feasibility_limit}.

\subsection{Baseline Optimization Problem}
\label{sub:baseline_problem}
The baseline optimization problem (P1) is formalized~as
\begin{subequations} \label{OP1}
	\begin{alignat}{2} 
	\text{P1}: \: 	&\!\min_{N_{\rm gNB}, P_t} \quad &E_C= \, &S_D P_{\rm gNB}T_{\rm SSB}, \tag{\ref*{OP1}} \\
		& & C_1: & \: P_t\gamma_k\geq \tau, \: \forall k, \nonumber \\
     & & C_2:& \: 0<P_t\leq P_T, \nonumber \\
		& & C_3:& \: N_{\rm gNB}\in \{i \in \mathbb{N} \; | \; i \leq 64\}. \nonumber
	\end{alignat}
\end{subequations}
Constraint $C_1$ ensures that the transmission power $P_t$ at the gNB is such that the SNR of each user is greater than or equal to a given threshold $\tau$, so \glspl{ue} can be properly detected. $C_2$ stipulates that $P_t$ be less than or equal to the maximum transmission power $P_T$, whereas $C_3$ sets the upper bound of the number of antennas at the gNB to $64$, in line with RedCap design guidelines~\cite{pagin2023nrlight}.

%It follows that the optimization problem determines the optimal value of $N_{\text{gNB}}$ and $P_t$, referred to as $N^*$ and $P_t^*$ such that the SNR$_k~\forall k$ is greater than a threshold $\tau$. From \eqref{gamma_k}, $\gamma_k$ depends on $G_{\rm gNB}$, and hence on the angular offset ${\theta}_k$ introduced by the moving UEs. As the UE moves at constant velocity $v$ during the beam management process, it may lose alignment with respect to the associated beam, potentially deteriorating the beamforming gain. This may cause the SNR of UE$_k$ to drop below the sensitivity threshold $\tau$, preventing it from being detected. The factors that may lead to misalignment include:
%(i) the UE velocity $v$ (the faster the UE, the sooner it may lose alignment); (ii) the beam management time $T_{\rm BM}$ and, consequently $T_{\rm SS}$ and $N_{\rm SS}$ (the slower the beam management procedure, the higher the probability that the UE would lose alignment); and (iii) the number of antennas $N_{\rm gNB}$, which defines the beamwidth (the narrower the beam, the higher the probability that the UE would lose alignment). In this work, we investigate the impact of these factors on the optimization problem.

%MP We are kind of repeating the same concept over and over, I'd consider removing part of this ?
Basically, the above optimization problem determines the optimal values of $N_{\text{gNB}}$ and $P_t$, referred to as $N^*_{\text{gNB}}$ and $P_t^*$, such that $E_C$ is minimized, and the SNR of each UE is greater than a threshold $\tau$. From Eq.~\eqref{gamma_k}, $\gamma_k$ depends on $G_{\rm gNB}$, and hence on the angular offset ${\theta}_k$ between the UEs and the boresight of the corresponding serving beam. 
%As the latter move throughout the beam management process, they may lose alignment with respect to the associated beam and potentially deteriorate their beamforming gain. %This may cause the SNR of UE$_k$ to drop below the sensitivity threshold $\tau$, preventing it from a detected. 
In turn, the misalignment is a function of:
(i) the UE velocity $v$ (the faster the \gls{ue}, the sooner it loses alignment); 
(ii) the beam management time $T_{\rm BM}$ and, consequently $T_{\rm SS}$ and $N_{\rm SS}$ (the slower the beam management procedure, the higher the probability that a \gls{ue} loses alignment); and 
(iii) the number of antennas $N_{\rm gNB}$, which in turn determines the beamwidth (the narrower the beam, the smaller the angular offset, which leads to a beamforming gain degradation). In the remainder of this work, we investigate the impact of these parameters on the above optimization problem.
%\smallskip 

%Consequently, the cumulative angular offset initially decreases and then increases with $N$ and the magnitude of the threshold or minimum offset decreases with $v$ and $T_{SS}$ and increases with $N_{SS}$. \

\subsection{Generalized Optimization Problem: Non-Zero Misdetection}
\label{sub:non-zero}
In Eq.~\eqref{OP1}, $C_1$ enforces that no UE violates the \gls{snr} constraint. 
To further generalize the problem, we also consider the case where some UEs are allowed to violate $C_1$ even though, statistically, most of the \glspl{ue} experience a sufficient link quality.
To this end, we introduce the misdetection probability $P_{\rm MD}$, and define an additional optimization problem which optimizes $N_{\rm gNB}$ and $P_t$ while ensuring that \textit{at least} $(1 -P_{\rm MD}) K$ users, out of $K$ \glspl{ue}, experience a sufficient \gls{snr}. 

The generalized problem (P2) can be formalized as
\begin{subequations} \label{OP2}
	\begin{alignat}{2}
	\text{P2}: \:	&\!\min_{N_{\rm gNB}, P_t} \quad       &E_C=&S_D P_{\rm gNB}T_{\rm SSB}, \tag{\ref*{OP2}} \\
		& & C_1^{'}: &\: \mathbb{P}_k [P_t\gamma_k\geq \tau]\geq 1-P_{\rm MD}, \nonumber \\
        & & C_2:& \: 0<P_t\leq P_T, \nonumber \\
		& & C_3:& \: N_{\rm gNB}\in \{i \in \mathbb{N} \; | \; i \leq 64\}. \nonumber 
	\end{alignat}
\end{subequations}
where $C_1^{'}$ ensures that the probability that UE$_k$ satisfies the \gls{snr} constraint is greater than or equal to $1-P_{\rm MD}$. In other words, $C_1^{'}$ allows the problem to be feasible even if $K P_{\rm MD}$ UEs lose alignment with the associated beam. 

\subsection{Optimization Algorithm}
\label{sec:opt_alg}
Due to the multiple nonlinearities in P1, such as the nested ceiling function and the dependence of $C_1$ on the integer $N_{\rm gNB}$, the optimization problem in Eq.~\eqref{OP1} is a \gls{minlp}, and thus non-convex. In fact, even powerful conventional solvers such as MATLAB's \texttt{surrogateopt}, and GAMS' ANTIGONE~\cite{misener2014antigone} and GUROBI~\cite{gurobi} failed to solve it. Therefore, we develop and present an \gls{mc} algorithm, reported in Alg.~\ref{algo1}, for obtaining the optimal solution of P1.

%Based on the optimization problem in~\eqref{OP1} and the considerations above, we conclude that there are two optimization variables, i.e., $P_t$ (transmit power at the gNB) and $N_{\rm gNB}$ (number of antenna elements at the gNB). 
%According to the expression of $E_C$ in Eq.~\eqref{EC1}, we observe that $P_{\rm PA}=P_t/\eta$ and hence $E_C$ increases linearly with $P_t$ and quadratically with $N_{\rm gNB}$. 
The algorithm is designed with the following rationale.
First, it can be observed that Eq.~\eqref{EC1} can be lower bounded as 
\begin{equation}
\label{EC_lb}   
\eqref{EC1} \geq 3 T_{\rm SSB} N_{\rm gNB} \left( {P_t}/{\eta} + N_{\rm gNB} P_{\rm PS} + P_{\rm RF}^{'} \right),
\end{equation}
where $P_{\rm RF}^{'} = P_{\rm RF} + P_{\rm LO} + 2P_{\rm LDAC}$. So, the derivatives of $E_C$ with respect to $P_t$ and $N_{\rm gNB}$ are non-negative, i.e., 
\begin{equation}
\label{EC_deriv_wrt_N}   
\frac{\partial E_C}{\partial N_{\rm gNB}} \geq 3T_{\rm SSB} \left( \frac{P_t}{\eta} + 2 N_{\rm gNB} P_{\rm PS} + P_{\rm RF}^{'} \right) > 0,
\end{equation}
\begin{equation}
\label{EC_deriv_wrt_PT}   
\frac{\partial E_C}{\partial P_{t}} \geq\frac{ 3 T_{\rm SSB} N_{\rm gNB}}{\eta} > 0,
\end{equation}
since feasible values of $N_{\rm gNB}$ are strictly positive (as per $C_3$).
Therefore, it can be concluded that the objective function drives the optimization problem to choose the minimum values of $N_{\rm gNB}$ and $P_t$ that meet the \gls{snr} constraint. 
%By expressions, $N_{\rm gNB}$ seems to be the dominant variable in this optimization problem.

By further inspecting Eqs.~\eqref{EC1},~\eqref{eq:SD}, and~\eqref{power_gNB}, we see that $E_C$ exhibits a quadratic dependence with respect to $N_{\rm gNB}$, and a linear dependence with respect to $P_t$. Therefore, we infer that $N_{\rm gNB}$ is the asymptotically dominant variable in the optimization problem.\footnote{The dominance of $N_{\rm gNB}$ with respect to $P_t$ has been verified experimentally for practical values of the problem parameters. }
This suggests that the minimum value of $N_{\rm gNB}$ for which $C_1$ is satisfied, assuming $P_t=P_T$, is the optimal $N_{\rm gNB}$, i.e., $N^*_{\rm gNB}$. 
Then, once $N^*_{\rm gNB}$ is fixed, the optimal transmission power $P_t^*$ can be computed as the minimum value which satisfies $C_1$ for $N_{\rm gNB} = N^*_{\rm gNB}$, i.e., $P_t^*=\tau/\min_k \gamma_k$.
Finally, whenever for a given value of $P_T$ and $\tau$ the constraints cannot be met with any feasible value of $N_{\rm gNB}$, the problem is assumed to be infeasible.

To generalize the algorithm for solving P2, we simply modify the condition of Line~\ref{alg:c1} in Alg.~\ref{algo1} as
\begin{equation}
\sum\limits_{k=1}^K \mathbbm{1}(P_t \gamma_k \geq \tau) \geq ( 1- P_{\rm MD}) K ,
\end{equation}
and the condition of Line~\ref{alg:pt} as 
\begin{equation}
P_t^* \leftarrow \tau / \left[ \bm{\gamma} \right]_{ \lceil ( 1- P_{\rm MD}) K \rceil  },
\end{equation}
where $\bm{\gamma}$ is the vector of the \gls{snr} values $\gamma_k$, sorted in descending order, and $\left[ \bm{\gamma} \right]_{i}$ is the $i$-th entry of $\bm{\gamma}$.

%Optimization algorithm: Compute all the feasible values of $N$, and choose the combination of ($N$, $P_t$) that gives the lowest energy consumption ($E_C$). 
%\SetKwComment{Comment}{/*}{}
\begin{algorithm}[t!]
	\caption{Computation of the solution of Eq.~\eqref{OP1}.}\label{algo1}
\small
	Initialize $P_T$, $\tau$, $K$, $R$;\\
	\For{each iteration $j = 1, \ldots, N_{\rm MC}$}{ 
	{ 
    \For{each \gls{ue} $k=1, \ldots, K$}{
		$\phi_k \gets 2\pi\text{rand}(1)$\; 
		$d_k \gets R\sqrt{\text{rand}(1)}$\; %\phi_k\in\phi, d_k\in d$ \Comment*[r]{user distribution parameters}
    }
		\For{$N_{\rm gNB}=1, 2, \dots, 64$}
    {
        $S_D=\left\lceil\pi N_{\rm gNB}\right\rceil$, $\Delta_{\rm 3dB}=2/N_{\rm gNB}$\;
        \For{each \gls{ue} $k=1, \ldots, K$}
        {
            %Compute $\theta_{i,k}$ using \eqref{theta_ik}, $\theta_{v,k}$ using \eqref{theta_vk}, $\theta_k$ using \eqref{theta_k}, and $\gamma_k$ using \eqref{gamma_k}\;
            Compute $\gamma_k$, $\theta_{i,k}$, $\theta_{v,k}$, and $\theta_k$, using Eqs.~\eqref{gamma_k},~\eqref{theta_ik},~\eqref{theta_vk}, and~\eqref{theta_k}, respectively\;
        }
        \If {$P_T \gamma_k\geq\tau \; \forall$ \gls{ue} $k$ \label{alg:c1} } 
        { $N^*_{j} \gets N_{\rm gNB}$\; 
        $P^*_{t, j}\gets\tau/\min_k{\gamma_k}$\; \label{alg:pt}
        break\;
		  }
	  }
	}
}
$N^*_{\rm gNB}=\sum_{i=j}^{N_{\rm MC}} N^*_{j}/N_{\rm MC}$\; \label{alg:n_gnb} %\text{MC}$\Comment*[r]{Monte Carlo summation}
$P_t^*=\sum_{i=j}^{N_{\rm MC}} P^*_{t, j}/N_{\rm MC}$\;
\end{algorithm}

\begin{figure*}[!t]
%\caption{Analytical expression for $\bar{\theta}$}
\centering
\begin{equation}
    \bar{\theta} =\int_{0}^{R} \int_{-\frac{\Delta_{3 \rm dB}}{2}}^{\frac{\Delta_{3 \rm dB}}{2}} \left\vert x + \frac{v T_{\rm SS}\left(\left\lceil \lceil \pi N_{g \rm NB} \rceil  /{N_{\rm SS}}\right\rceil-1\right)+T_{\ell} }{y} \right\vert \frac{2y}{R^2} \frac{N_{g \rm NB}}{2} dx \, dy. \label{an_long}
\end{equation}
\vspace{0.1cm}
\noindent\rule{\textwidth}{0.2pt}
\vspace{-0.5cm}
\end{figure*}

\section{Optimization Metrics}
Based on the optimization problem in Sec.~\ref{sec3}, we introduce and formalize two fundamental optimization metrics that will be used to evaluate and dimension RedCap beam management, namely the average angular offset (Sec.~\ref{antenna_Gain}) and the feasibility regions (Sec.~\ref{feasibility_limit}).

\subsection{Angular Offset} \label{antenna_Gain}

The first optimization metric that we consider for evaluating the solutions of P1 and P2 is the average angular offset $\bar{\theta}$ of all $K$ UEs in the region.
%, for which we derive analytical expressions.
Since users are uniformly distributed in the circular area, the probability density functions of the initial phase $\phi_k$ and distance $d_k$ relative to the center are respectively written as
\begin{equation}
    p_{\phi_k}(x) =
    \begin{cases}
    \frac{1}{2 \pi} &\text{if $x \in [ 0, 2 \pi ) $ }\\
    0 &\text{otherwise,}
    \end{cases}  
    % \:  p_{d_k}(x) =
    %  \begin{cases}
    % \frac{2x}{R^2} &\text{if $x \in [0, R] $ }\\
    % 0 &\text{otherwise}.
\end{equation}
% and
\begin{equation}
    p_{d_k}(x) =
    \begin{cases}
    \frac{2x}{R^2} &\text{if $x \in [0, R] $ }\\
    0 &\text{otherwise}.
    \end{cases}  
\end{equation}

For a given number of antennas $N_{\rm gNB}$, the serving region is subdivided into $S_D$ regions of equal area. %which span the whole coverage area. 
Since $\phi_k$ is uniformly distributed, UE$_k$ is under the initial coverage of beam $m\in\{1,\dots,S_D\}$ with probability $\mathbb{P}(m)=1/S_D$.
 %the probability of ending up in the $m$-th beam coverage area is ${1}/{S_D}$, which is independent of $m$. 
Therefore, the probability density function of the initial offset $\theta_{i, k}$ with respect to the boresight direction of beam $m$ is
\begin{equation*}
    p_{\theta_{i,k}}(x \mid m)=
    \begin{cases}
    {1}/{\Delta_{3 \rm dB}} &\text{if $x \in [ {-\Delta_{3 \rm dB}}/{2}, {\Delta_{3 \rm dB}}/{2} ] $ }\\
    0 &\text{otherwise.}
    \end{cases}  
\end{equation*}

Then, the average angular offset $\bar{\theta}$ can be expressed as
\begin{align}
    \bar{\theta} =& \sum_{m=1}^{S_D} \mathbb{E} [\theta \mid m] \mathbb{P}(m) \\
    =&  \sum_{m=1}^{S_D} \mathbb{E} [\theta \mid m] ({1}/{S_D}) = \mathbb{E} [\theta \, | \, m],
\end{align}
so  $\bar{\theta}$ does not depend on the specific beam $m$.
Therefore, $\bar{\theta}$ can be evaluated by focusing on the coverage region of a single beam as
\begin{align}
    \bar{\theta} &= \mathbb{E} [\lvert  \theta_{i, k} + \theta_{v, k} \rvert \mid m ] \\ 
    % &= \mathbb{E} [\theta \, | \, m] \\
    &= \int_{0}^{R} \int_{-\frac{\Delta_{3 \rm dB}}{2}}^{\frac{\Delta_{3 \rm dB}}{2}} \left\vert \theta_{i, k} + \theta_{v, k} \right\vert p(x)_{\theta_{i, k}} p(y)_{d_k} dx \, dy \\
    &= \int_{0}^{R} \int_{-\frac{\Delta_{3 \rm dB}}{2}}^{\frac{\Delta_{3 \rm dB}}{2}} \left\vert x + \frac{v T_{\rm BM}}{y} \right\vert \frac{2y}{R^2} \frac{1}{\Delta_{3 \rm dB}} dx \, dy. \label{an_short}
\end{align}

Rendering explicit the dependence on $N_{g \rm NB}, N_{\rm SS}$ and $T_{\rm SS}$, Eq.~\eqref{an_short} can be further manipulated into Eq.~\eqref{an_long}.

\subsection{Feasibility Regions} \label{feasibility_limit}
Given the maximum transmission power $P_T$, the \gls{snr} threshold $\tau$, and the number of users $K$, P1 and P2 yield the optimal number of antennas $N^*_{\rm gNB}$ and transmission power $P_t^*$ at the gNB.
Based on these values, we determine the upper limits of the ground speed $v$ and the burst periodicity $T_{\rm SS}$ which prevent the misalignment of the \glspl{ue} with respect to the corresponding associated beam. 
To this end, we define the alignment condition of UE$_k$, with initial (accumulated) offset $\theta_{i,k}$ ($\theta_{v,k}$), as being located within the angular coverage region of its associated beam. This can be formalized as
\begin{align}
	\theta_{i,k}+\theta_{v,k} \leq \frac{\Delta_{\rm FNBW}}{2},
 \label{offset}
\end{align}
where $\Delta_{\rm FNBW}=2 \sin^{-1}\left[2/N^*_{\rm gNB}\right]$ is the \gls{fnbw}. Therefore, Eq.~\eqref{offset} bounds the maximum angular offset to the first null beamwidth, i.e., the angular region between the first zeros of the main lobe.

Given the independence of $\theta_{i,k}$ and $\theta_{v,k}$, and substituting the expression of $\theta_{v,k}$ from Eq.~\eqref{theta_vk}, we get
\begin{align}\nonumber
	\theta_{i,k} + \frac{vT_{SS}}{d_k}\left(\left\lceil\frac{S_D}{N_{SS}}\right\rceil-1\right)\leq \frac{\Delta_{\rm FNBW}}{2},
\end{align}
which can be further manipulated into
\begin{align}
	vT_{SS}\leq d_k\left[\dfrac{\Delta_{\rm FNBW}/2- \theta_{i,k}}{\left(\left\lceil S_D/N_{\rm SS}\right\rceil-1\right)}\right],
 \label{vTss_lim}
\end{align}
thus providing an upper limit for the feasible product of $v$ and $T_{\rm SS}$ for given values of $N^*_{\rm gNB}$, $N_{\rm SS}$, and $K$.
Finally, Eq.~\eqref{vTss_lim} can be evaluated by substituting the random quantities $d_k$ and $\theta_{i,k}$ with their deterministic worst-case counterparts, and taking the value of $\Delta_{\rm FNBW}$ which corresponds to the highest possible gain, i.e.,
\begin{align}
	vT_{SS}\leq \min_k d_k\left[\dfrac{\Delta_{\rm FNBW}/2-\max_k \theta_{i,k}}{\left(\left\lceil S_D/N_{\rm SS}\right\rceil-1\right)}\right].
 \label{vTss_lim_det}
\end{align}
%This has been used to produce an upper bound on the feasible limits obtained through simulations in Sec.~\ref{sec_FL}.
%the feasibility of $v, T_{\rm SS}$, and $N_{\rm SS}$ for a given $N^*$ is limited by the equation (assuming $P_{MD}=0$)

% \subsection{Average Power Consumption}
% While in P1 and P2 we minimize the energy consumption of beam management, we now evaluate the average power consumption for sending \glspl{ssb} over time. This is defined as the cumulative energy consumed for sending \glspl{ssb} during a complete beam management sweep divided by its duration, so it is itself a function of $N_{\rm gNB}$. It can be computed as
% \begin{align}
% 	P_C=P_{\rm gNB}T_{\rm SSB}{N_{\rm SS}}/{T_{\rm SS}},
%  \label{ECT}
% \end{align}
% where $P_{\rm gNB}T_{\rm SSB}$ represents the energy consumed for sending one SSB as per Eq.~\eqref{EC1}. 

%In the following section, we leverage this metric to propose the optimal beam management parameters (in an energy efficiency sense) for a given gNB configuration. %\hl{$P_C$ has a linear dependence on both $N_{\rm gNB}$ and $P_t$ and can be a better parameter in analyzing power consumption in a frame structure.}

\begin{table}[t!]
\centering
\scriptsize
\renewcommand{\arraystretch}{0.95}
\caption{Simulation parameters.}
\label{parameter}
	\begin{tabular}{ l|l|l }
		\hline
		{Parameter} & {Description} & {Value}\\
		\hline
		$h_{\rm gNB}$&  gNB height  & $35$ m\\
		$h_{\rm UE}$ & UE height & $1.5$ m\\
        %$h$ & Average building height& $5$ m\\
        %$W$ & Average street width & $20$ m\\
		$R$ &  Radius of the SmA scenario & $100$ m \\ %\hline
		$K$   & Number of UEs  & $\{50, 100, 200, 500, 1000\}$\\
        $f_c$   & Carrier frequency  & $28$ GHz\\
        $B$ & Bandwidth  & $50$ MHz\\
        % $P_T$ & Transmission power  & $18$ dBm\\
        %$\tau$ & SNR threshold  & $\{3, 7, 10\}$ dB\\
		$\mathbbm{h}_{\rm LoS}^{k}, \mathbbm{h}_{\rm NLoS}^{k}$ & Channel fading gains & $\mathcal{CN}(0,1)$ \\
        $\text{PL}_{\rm LoS}^k$, $\text{PL}_{\rm NLoS}^k$ & Path loss & \cite[Table 7.4.1-1]{3GPPrel16}\\
        $P_r(d_{3D})$ & LoS probability & \cite[Table 7.4.2-1]{3GPPrel16} \\
        $N_0$ & Noise \gls{psd}  & $-174$ dBm/Hz\\
        $n$ & 5G NR numerology index & $4$ \\
        $N_{\rm MC}$ & Number of \gls{mc} simulations & $10^5$\\ \hline
        $\eta$ & Power added efficiency & $27\%$ \\
        $b_{\rm DAC}$ & DAC bit resolution & $8$ \\
        $F_s$ & DAC sampling frequency & $10^9$ Hz \\\hline
        $T_{\rm SS}$ & SS burst period & $\{5, 10, 20, 40 ,80, 160\}$~ms\\
        $N_{\rm SS}$ & Number of SSBs per burst & $\{8, 16, 32, 64\}$\\
        $v$ & \gls{ue} speed & $[1,30]$ m/s\\
        $P_T$ & Maximum transmission power & $[10,40]$ dBm\\
        $\tau$ & \gls{snr} threshold & $[1,10]$ dB\\\hline
       % $N^*, P_t^*$ & optimal number of antenna elements at gNB & 
        \end{tabular}
\end{table}
%\vspace{-0.5em}

\begin{figure}
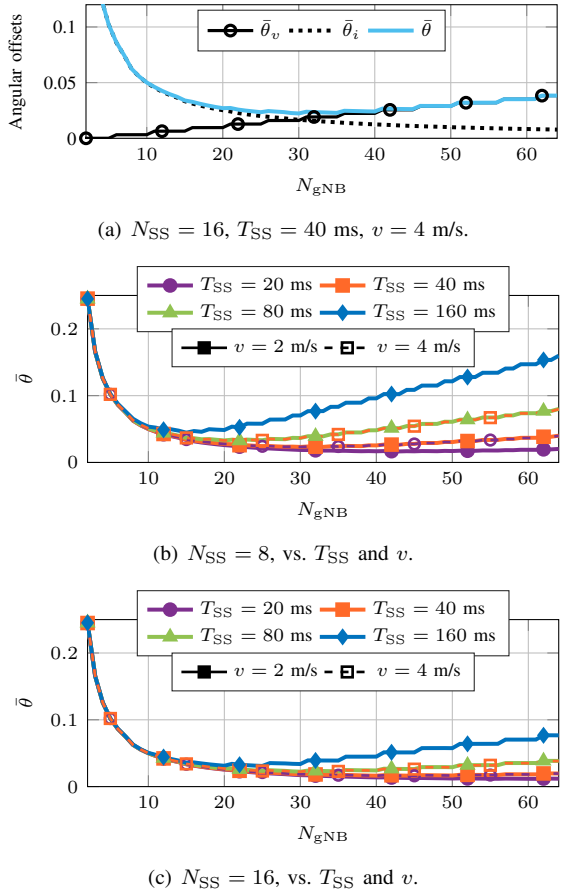

	\centering 
	\setlength\fwidth{0.7\columnwidth}
    \setlength\fheight{0.2\columnwidth}
	\subfigure[$N_{\rm SS}=16$, $T_{\rm SS}=40$ ms, $v=4$ m/s.] {\definecolor{mycolor1}{rgb}{0.92900,0.69400,0.12500}%
\definecolor{mycolor2}{rgb}{1.00000,0.07451,0.65098}%
\definecolor{mycolor3}{rgb}{0.46600,0.67400,0.18800}%
\definecolor{mycolor4}{rgb}{0.30100,0.74500,0.93300}%
\definecolor{mycolor4}{rgb}{0.30100,0.74500,0.93300}%
\definecolor{mycolor5}{rgb}{0.30196,0.74510,0.93333}%
\begin{tikzpicture}

\begin{axis}[%
width=\fwidth,
height=\fheight,
at={(0.531\fwidth,0.5\fheight)},
scale only axis,
xmin=2,
xmax=64,
ymin=0,
ymax=0.12,
axis background/.style={fill=white},
title style={font=\bfseries},
xlabel={$N_{\rm gNB}$},
ylabel={Angular offsets},
ytick={0, 0.05, 0.1},
yticklabels={0, 0.05, 0.1},
xmajorgrids,
ymajorgrids,
legend style={at={(0.5,0.65)}, anchor=south, legend columns=3, legend cell align=left, align=left, draw=white!15!black}
]

\addplot [color=black, mark=o, line width=1.2pt, mark repeat=10]
  table[row sep=crcr]{%
  1	9.26742493327765e-06\\
  2	1.71052923134491e-05\\
  3	2.42366139710383e-05\\
  4	3.20744813512098e-05\\
  5	3.9205803008799e-05\\
  6	0.00320055282767722\\
  7	0.00320768414933481\\
  8	0.00321766360869332\\
  9	0.00322550147607349\\
  10	0.00323263279773108\\
  11	0.0063939798223995\\
  12	0.00640111114405709\\
  13	0.00640894901143726\\
  14	0.00641608033309485\\
  15	0.00642605979245336\\
  16	0.00958740681712178\\
  17	0.00959453813877938\\
  18	0.00960237600615955\\
  19	0.00960950732781714\\
  20	0.00961734519519731\\
  21	0.0127779856741431\\
  22	0.0127879651335017\\
  23	0.0127958030008818\\
  24	0.0128029343225394\\
  25	0.0128107721899196\\
  26	0.0159714126688654\\
  27	0.0159792505362456\\
  28	0.0159863818579032\\
  29	0.0159963613172617\\
  30	0.0160041991846419\\
  31	0.0191648396635877\\
  32	0.0191726775309679\\
  33	0.0191798088526255\\
  34	0.0191876467200057\\
  35	0.0191947780416632\\
  36	0.02235826665831\\
  37	0.0223661045256902\\
  38	0.0223732358473477\\
  39	0.0223810737147279\\
  40	0.0223882050363855\\
  41	0.0255495520610539\\
  42	0.0255566833827115\\
  43	0.02556666284207\\
  44	0.0255745007094502\\
  45	0.0255816320311078\\
  46	0.0287429790557762\\
  47	0.0287501103774338\\
  48	0.028757948244814\\
  49	0.0287650795664716\\
  50	0.0287750590258301\\
  51	0.0319364060504985\\
  52	0.0319435373721561\\
  53	0.0319513752395363\\
  54	0.0319585065611939\\
  55	0.031966344428574\\
  56	0.0319734757502316\\
  57	0.0351369643668784\\
  58	0.0351448022342585\\
  59	0.0351519335559161\\
  60	0.0351597714232963\\
  61	0.0351669027449539\\
  62	0.0383282497696223\\
  63	0.0383353810912799\\
  64	0.0383453605506384\\
};
\addlegendentry{$\bar{\theta}_v$}

\addplot [color=black, dotted, line width=1.5pt]
  table[row sep=crcr]{%
  1	0.484205182991397\\
  2	0.245125368041322\\
  3	0.165622092601125\\
  4	0.125704406033933\\
  5	0.101892710306382\\
  6	0.0859251934192471\\
  7	0.0745949359429686\\
  8	0.062306433915695\\
  9	0.0553478014053999\\
  10	0.0499497103089043\\
  11	0.0455428611126488\\
  12	0.041977475323159\\
  13	0.0389840822086757\\
  14	0.0364753170120083\\
  15	0.0333163958574434\\
  16	0.0312008951755617\\
  17	0.0294232554031564\\
  18	0.0278276558701831\\
  19	0.0264332404045142\\
  20	0.0252341077571605\\
  21	0.02413268503179\\
  22	0.022702689526516\\
  23	0.0216953215440175\\
  24	0.0208311911169541\\
  25	0.0200022480858016\\
  26	0.0192879833425873\\
  27	0.018661809937249\\
  28	0.0180622100966874\\
  29	0.0172396237990756\\
  30	0.0166557923314942\\
  31	0.0161413377314246\\
  32	0.0156496302514382\\
  33	0.0151853702880308\\
  34	0.0147794029711474\\
  35	0.0143945514899534\\
  36	0.013871058677315\\
  37	0.0135086611591035\\
  38	0.013167054123074\\
  39	0.0128182635115081\\
  40	0.0125149758797333\\
  41	0.0122505897055423\\
  42	0.0119702606833341\\
  43	0.0116262405170545\\
  44	0.0113460508803591\\
  45	0.0111022163408415\\
  46	0.0108834177069448\\
  47	0.0106389356709827\\
  48	0.0104521947809777\\
  49	0.0102615610073962\\
  50	0.00997749745594206\\
  51	0.00979477542835479\\
  52	0.00961791571938988\\
  53	0.0094343622129708\\
  54	0.00926610265429331\\
  55	0.00910385371234202\\
  56	0.00896441916102539\\
  57	0.00876922339602119\\
  58	0.00861643874296397\\
  59	0.00847775177600417\\
  60	0.00834641887802228\\
  61	0.00820427327819615\\
  62	0.00809071449987532\\
  63	0.00796817187485927\\
  64	0.007818277230937\\
};
\addlegendentry{$\bar{\theta}_i$}

\addplot [color=mycolor5, line width=1.5pt]
  table[row sep=crcr]{%
  1	0.484205574601604\\
  2	0.245126106754323\\
  3	0.165623307281601\\
  4	0.125705880097007\\
  5	0.101894590219359\\
  6	0.0861469118169613\\
  7	0.0748389734523177\\
  8	0.0624352891989972\\
  9	0.0555261867685641\\
  10	0.0501389863947583\\
  11	0.0461725058950307\\
  12	0.0426596441796622\\
  13	0.0397006677578307\\
  14	0.0372505442408202\\
  15	0.0339967422756688\\
  16	0.0327373648581379\\
  17	0.0310228595907254\\
  18	0.0295096635318811\\
  19	0.0281912150615011\\
  20	0.0270798926923092\\
  21	0.0272620495034141\\
  22	0.0258146949816478\\
  23	0.024912472596294\\
  24	0.0241826194211825\\
  25	0.0234611763564966\\
  26	0.0244923808512652\\
  27	0.0239930729270036\\
  28	0.0235509102995268\\
  29	0.0227078357396051\\
  30	0.0223075455732391\\
  31	0.0238856892207304\\
  32	0.0235583826339739\\
  33	0.0232996177869598\\
  34	0.0230869151083608\\
  35	0.0228341395714709\\
  36	0.0247313706428649\\
  37	0.0245524947955802\\
  38	0.0243880987481868\\
  39	0.0242042207125126\\
  40	0.0240983915580377\\
  41	0.0266609848828902\\
  42	0.0265967993551476\\
  43	0.0263762966315755\\
  44	0.0262928135755146\\
  45	0.0262196474069646\\
  46	0.0291076835039459\\
  47	0.0290661597396396\\
  48	0.0290707835049001\\
  49	0.0290516281219427\\
  50	0.0289362331935194\\
  51	0.0319756027863793\\
  52	0.0320018418925379\\
  53	0.0319863212802029\\
  54	0.0320070134137718\\
  55	0.0320135309530189\\
  56	0.032056728820943\\
  57	0.0351389304335793\\
  58	0.0351416817915824\\
  59	0.0351292549601132\\
  60	0.0351739199461374\\
  61	0.0352187693005817\\
  62	0.0383740370021409\\
  63	0.0383955359633275\\
  64	0.0383427465734795\\
};
\addlegendentry{$\bar{\theta}$}

\end{axis}
\end{tikzpicture}\label{theta_components}}
	\centering
    \setlength\fwidth{0.7\columnwidth}
    \setlength\fheight{0.25\columnwidth}
	\subfigure[$N_{\rm SS}=8$, vs. $T_{\rm SS}$ and $v$.] {\input{Figures/theta_bar_Nss8}\label{theta_bar_Nss8}}
    \centering 
    \setlength\fwidth{0.7\columnwidth}
    \setlength\fheight{0.25\columnwidth}
	\subfigure[$N_{\rm SS}=16$, vs. $T_{\rm SS}$ and $v$.] {\input{Figures/theta_bar_Nss16}\label{theta_bar_Nss16}}
    \caption{Average angular offset $\bar{\theta}$ as a function of $N_{\rm gNB}$, $v$, and $T_{\rm SS}$. In the two bottom figures, solid lines and filled markers represent $v=2$ m/s, while dashed lines and empty markers represent $v=4$ m/s.}\label{theta_bar}
\end{figure}

\section{Numerical Results}\label{results}
In this section we numerically evaluate the optimal number of antennas $N^*_{\rm gNB}$ and transmission power $P_t^*$ to minimize $E_C$ during beam management in different system configurations, for $N_{\rm MC} = 10^5$ Monte Carlo simulations. We study the impact of the beam management parameters, the maximum transmission power, the SNR threshold, and the number of users on the solution of P1, and compare it with the solution of P2 when the misdetection probability is not zero.
Moreover, we evaluate the feasible angular offsets, and %consider the case of non-zero misdetection probabilities. %on $N^*_{\rm gNB}$ and $P_t^*$ and the feasibility limits, and the trade-off. 
%Finally, we 
define the feasibility regions for RedCap devices, which strike an optimal trade-off between \gls{qos} and $E_C$.

In our simulations, we used the 3GPP LoS probability and path loss parameters for \gls{rma} scenarios reported in~\cite[Tables 7.4.1-1, 7.4.2-1]{3GPPrel16}. The parameters for the \glspl{ue} and gNB are from the RedCap design guidelines~\cite{pagin2023nrlight}, while the power consumption parameters are taken from~\cite{energyefficiency2018}, as summarized in Table~\ref{parameter}.
%All the numerical results have been obtained for . 

\subsection{Angular Offset} \label{modeling}

In Fig. \ref{theta_bar} we plot the angular offset $\bar{\theta}$ averaged over all $K$ UEs in the region, versus $N_{\rm gNB}$, and for different values of $v$, $N_{\rm SS}$ and $T_{\rm SS}$. 
Notice that $\bar{\theta}$ depends on the average initial offset $\bar{\theta}_{i}$ and the average offset due to mobility $\bar{\theta}_{v}$. 
As depicted in Fig.~\ref{theta_components}, the former decreases monotonically with $N_{\rm gNB}$ since the beamwidth $\Delta_{\rm 3dB}$ and hence the angular distance between consecutive boresight directions decrease. Conversely, the latter increases (through not monotonically due to the ceiling function in Eq.~\eqref{theta_vk}) with $N_{\rm gNB}$ since the number of SSBs required for a complete beam scan increases. Overall, $\bar{\theta}$ decreases and then increases with $N_{\rm gNB}$.

When the number of antennas is small, e.g., for $N_{\rm gNB} \leq 10$, beams are sufficiently large to ensure continuous alignment of the \glspl{ue}, despite their mobility. In this case, $\bar{\theta}$ is dominated by the initial offset $\bar{\theta}_{i}$, so $\bar{\theta}$ decreases with $N_{\rm gNB}$ as reported in Fig.~\ref{theta_bar_Nss8}.
As $N_{\rm gNB}$ increases, the beams become progressively narrower, and the number of SSBs that are required to exhaustively scan the angular space increases. As such, the beam management time increases as per Eq.~\eqref{theta_vk}, so $\bar{\theta}$ is dominated by $\bar{\theta}_v$, and the impact of $v$ is non-negligible.

From Figs.~\ref{theta_bar_Nss8} and~\ref{theta_bar_Nss16} we further observe that $\bar{\theta}$ grows with $v$ and $T_{\rm SS}$ when $N_{\rm gNB}\geq 10$, even though this effect can be partially mitigated by increasing $N_{\rm SS}$. 
Additionally, we observe that the angular offset for $v=4$ m/s and $T_{\rm SS}=20$ ms is equal to the offset for $v=2$ m/s and $T_{\rm SS}=40$ ms. Similarly, the offset for $v=4$ m/s and $T_{\rm SS}=40$ is equal to the offset for $v=2$ m/s and $T_{\rm SS}=80$ ms. 
Therefore, we conclude that $\bar{\theta}$ depends on $v$ and $T_{\rm SS}$ only through their product, an observation which will be particularly relevant for the analysis of the feasibility regions in Sec.~\ref{sec_FL}.

\begin{figure}
	\centering 
    \setlength\fwidth{0.7\columnwidth}
    \setlength\fheight{0.3\columnwidth}
	\input{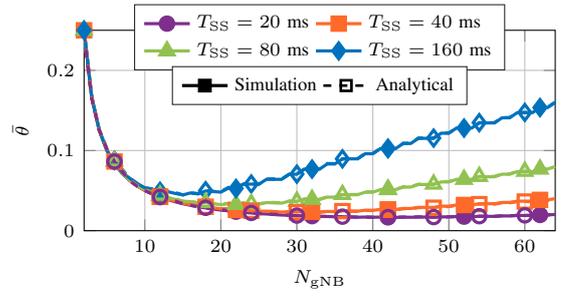}
    \label{theta_bar_Nss8_an_vs_sim}
	\caption{Average angular offset $\bar{\theta}$ obtained from Eq.~\eqref{an_long} (dashed lines and empty markers) and via Monte Carlo simulation (solid lines and filled markers), as a function of $N_{\rm gNB}$ and $T_{\rm SS}$, for $v=2$ m/s and $N_{\rm SS}=8$.}\label{theta_bar_an_vs_sim}
\end{figure}

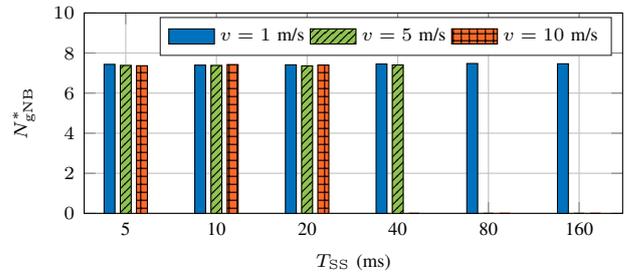
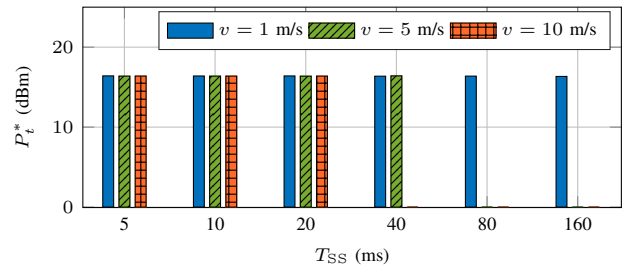
\begin{figure}[b]
	\centering 
    \setlength\fwidth{0.8\columnwidth}
    \setlength\fheight{0.3\columnwidth}
	\subfigure[Optimal number of antenna elements $N^*_{\rm gNB}$ at the gNB.] {% This file was created by matlab2tikz.
%
%The latest updates can be retrieved from
%  http://www.mathworks.com/matlabcentral/fileexchange/22022-matlab2tikz-matlab2tikz
%where you can also make suggestions and rate matlab2tikz.
%
\definecolor{mycolor1}{rgb}{0.00000,0.44700,0.74100}%
\definecolor{mycolor2}{rgb}{0.85098,0.32549,0.09804}%
\definecolor{mycolor3}{rgb}{0.92900,0.69400,0.12500}%
\definecolor{mycolor4}{rgb}{0.30196,0.74510,0.93333}%
\definecolor{mycolor5}{rgb}{0.56863,0.76078,0.31765}%
\definecolor{mycolor6}{rgb}{0.49412,0.18431,0.55686}%
\definecolor{mycolor7}{rgb}{1.00000,0.41176,0.16078}%
\begin{tikzpicture}

\begin{axis}[%
width=\fwidth,
height=\fheight,
at={(0\fwidth,0\fheight)},
scale only axis,
bar shift auto,
xmin=0.541423076923077,
xmax=6.48,
xtick={1,2,3,4,5,6},
xticklabels={{5},{10},{20},{40},{80},{160}},
xlabel={$T_{\rm SS}$ (ms)},
ymin=0,
ymax=10,
ylabel={$N^*_{\rm gNB}$},
axis background/.style={fill=white},
xmajorgrids,
ymajorgrids,
legend style={legend columns=3, legend cell align=left, align=left, draw=white!15!black}
]
\addplot[ybar, bar width=0.12, fill=mycolor1, draw=black, area legend] table[row sep=crcr] {%
1	7.4366\\
2	7.3972\\
3	7.4134\\
4	7.4514\\
5	7.4792\\
6	7.4644\\
};
\addlegendentry{$v=1$ m/s}

\addplot[ybar, bar width=0.12, preaction={fill, mycolor5}, pattern={north east lines},  draw=black, area legend] table[row sep=crcr] {%
1	7.3814\\
2	7.3808\\
3	7.3594\\
4	7.4058\\
5	0\\
6	0\\
};
\addlegendentry{$v=5$ m/s}

\addplot[ybar, bar width=0.12, preaction={fill, mycolor7}, pattern={grid}, draw=black, area legend] table[row sep=crcr] {%
1	7.3638\\
2	7.4204\\
3	7.3988\\
4	0\\
5	0\\
6	0\\
};
\addlegendentry{$v=10$ m/s}
\end{axis}

\end{tikzpicture}%
}\label{Nopt_Tss}
	\subfigure[Optimal transmission power $P_t^*$ at the gNB.] {% This file was created by matlab2tikz.
%
%The latest updates can be retrieved from
%  http://www.mathworks.com/matlabcentral/fileexchange/22022-matlab2tikz-matlab2tikz
%where you can also make suggestions and rate matlab2tikz.
%
\definecolor{mycolor1}{rgb}{0.00000,0.44700,0.74100}%
\definecolor{mycolor2}{rgb}{0.85000,0.32500,0.09800}%
\definecolor{mycolor3}{rgb}{0.92900,0.69400,0.12500}%
\definecolor{mycolor4}{rgb}{0.30196,0.74510,0.93333}%
\definecolor{mycolor5}{rgb}{0.46600,0.67400,0.18800}%
\definecolor{mycolor6}{rgb}{0.49412,0.18431,0.55686}%
\definecolor{mycolor7}{rgb}{1.00000,0.41176,0.16078}%
\begin{tikzpicture}

\begin{axis}[%
width=\fwidth,
height=\fheight,
at={(0\fwidth,0\fheight)},
scale only axis,
bar shift auto,
xmin=0.541423076923077,
xmax=6.48142307692308,
xtick={1,2,3,4,5,6},
xticklabels={{5},{10},{20},{40},{80},{160}},
xlabel={$T_{\rm SS}$ (ms)},
ymin=-0.0590403260150469,
ymax=25,
ylabel={$P_t^*$ (dBm)},
axis background/.style={fill=white},
xmajorgrids,
ymajorgrids,
legend style={legend columns=3, legend cell align=left, align=left, draw=white!15!black}
]
\addplot[ybar, bar width=0.12, fill=mycolor1, draw=black, area legend] table[row sep=crcr] {%
1	16.4072264724837\\
2	16.3948550463632\\
3	16.4005991341392\\
4	16.3700560377173\\
5	16.3778870435831\\
6	16.3383304365082\\
};
\addlegendentry{$v=1$ m/s}

\addplot[ybar, bar width=0.12, preaction={fill, mycolor5}, pattern={north east lines}, draw=black, area legend] table[row sep=crcr] {%
1	16.3749761586049\\
2	16.3768341940913\\
3	16.3713220335545\\
4	16.4051810907523\\
5	0\\
6	0\\
};
\addlegendentry{$v=5$ m/s}

\addplot[ybar, bar width=0.12, preaction={fill, mycolor7}, pattern={grid}, draw=black, area legend] table[row sep=crcr] {%
1	16.3837433152047\\
2	16.3781641832062\\
3	16.3786572448694\\
4	0\\
5	0\\
6	0\\
};
\addlegendentry{$v=10$ m/s}
\end{axis}

\end{tikzpicture}%
\label{Ptopt_Tss}}
	\caption{$N^*_{\rm gNB}$ and $P_t^*$ as a function of $v$ and $T_{\rm SS}$, for $N_{\rm SS}=8$, $P_T=18$ dBm, $\tau=7$ dB, and $K=50$. Missing values represent infeasibility.}
    \label{NPt_Tss}
\end{figure}

In Fig.~\ref{theta_bar_an_vs_sim} we also report the average angular offset obtained both via Monte Carlo simulations and by evaluating the analytical expression in Eq.~\eqref{an_long}. We can see that the curves perfectly match, thus validating our analytical results.

\begin{comment}
The results are obtained in following categories:  
\begin{enumerate}
    \item $N^*$ as a function of $T_{\rm SS}$, the UE velocity $v$, and number of SSBs per burst ($N_{\rm SS}$).
    \item $N^*$ as a function of the transmission power ($P_T$), the SNR threshold ($\tau$), and the number of UEs ($K$).
    \item Energy consumption ($E_C$) and energy consumption per unit of time (ECT).
\item Feasibility regions in terms of $N_{\rm SS}$ and $T_{\rm SS}$, and as a function of $K$.
\item Impact of the non-zero misdetection probability.
\item Trade-offs to identify the optimal beam management parameters for REDCAP devices.
\end{enumerate}
\end{comment}

%\begin{Definitions}{}{}
\Summary
The angular offset depends on the product of $v$ and $T_{\rm SS}$, and is non-monotonic with respect to $N_{\rm gNB}$. In particular, the angular offset is dominated by the initial angular offset when $N_{\rm gNB}$ is small, and then by the angular offset due to mobility when $N_{\rm gNB}$ increases.
%\end{Definitions}

\subsection{Impact of the Beam Management Parameters%such as $v$, $T_{\rm SS}$, and $N_{\rm SS}$
}\label{results1}

\begin{figure}
	\centering 
        \setlength\fwidth{0.75\columnwidth}
        \setlength\fheight{0.33\columnwidth}
	\input{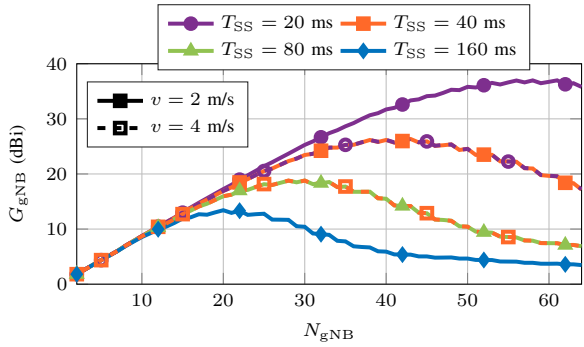}
	\caption{Average beamforming gain $G_{\rm gNB}$ at the \gls{gnb} as a function of $N_{\rm gNB}$, $v$ and~$T_{\rm SS}$, for $N_{\rm SS}=8$. Solid lines and filled markers represent $v=2$ m/s, while dashed lines and empty markers represent $v=4$ m/s.}
	\label{Gain_N}
\end{figure}

\begin{figure}[t!]
	\centering 
     \setlength\fwidth{0.75\columnwidth}
     \setlength\fheight{0.25\columnwidth}
	\input{Figures/PT_grouped}
	\caption{$N^*_{\rm gNB}, P_t^*, E_C$, and $P_C$ as a function of $P_T$ and $K$, for $v=1$ m/s, $T_{\rm SS}=20$ ms, $N_{\rm SS}=8$, and $\tau=-5$ dB.}
        \label{fig:PT_grouped}
\end{figure}

% \begin{figure}
% 	\centering 
%     \setlength\fwidth{0.75\columnwidth}
%     \setlength\fheight{0.27\columnwidth}
% 	\input{Figures/Nopt_Nss}
% 	\caption{$N^*_{\rm gNB}$ as a function of $N_{\rm SS}$ and $T_{\rm SS}$, for $P_T=18$ dBm, $\tau=7$ dB, and $v=5$ m/s. Missing values represent infeasibility.}
% 	\label{Nopt_Nss}
% \end{figure}

Fig.~\ref{NPt_Tss} depicts $N^*_{\rm gNB}$ and $P_t^*$ as a function of $T_{\rm SS}$ and $v$, for $N_{\rm SS}=8$, $P_T=18$ dBm, $\tau=7$ dB, and $K=50$.\footnote{We remark that, despite Constraint $C_3$ in P1, $N^*_{\rm gNB}$ is possibly non-integer, since it represents the average value across several Monte Carlo realizations (see Line~\ref{alg:n_gnb} in Alg.~\ref{algo1}).}  
We observe that the values of $N^*_{\rm gNB}$ and $P_t^*$ do not change much with respect to both $v$ and $T_{\rm SS}$ at $N_{\rm SS}=8$. This is motivated by the fact that the objective function drives the optimization problem towards the minimum values of $N^*_{\rm gNB}$ and $P_t^*$ that meet the SNR constraint, so as to minimize $E_C$. 
Therefore, $N^*_{\rm gNB}$ takes the minimum value corresponding to the largest offset beyond which the problem is infeasible, which also determines the corresponding values of $v$ and $T_{\rm SS}$.
In the given configuration, we obtain $N^*_{\rm gNB}=7.4$ and $P_t^*=16.4$ dBm.
Notice that some bars are missing in Fig.~\ref{NPt_Tss}, e.g., for $v\geq 5$ m/s when $T_{\rm SS} \geq 80$ ms and $N_{\rm SS}=8$, which indicates that the problem is infeasible.
That is to say, for given values of $P_T$ and $\tau$, only some values of $v$ and $T_{\rm SS}$ are feasible.

Moreover, as depicted in Fig.~\ref{Gain_N}, the beamforming gain $G_{\rm gNB}$ is approximately constant with respect to $v$ and $T_{\rm SS}$, as long as $N_{\rm gNB}$ is small (approximately, $N_{\rm gNB} < 10$). In this case, misalignment is mainly due to the initial offset (see Sec.~\ref{modeling}), which does not depend on $v$ or $T_{\rm SS}$.

\Summary
	The impact of $T_{\rm SS}$ and $N_{\rm SS}$ on $N^*_{\rm gNB}$ and $P_t^*$ is negligible. Still, for given values of $P_T$ and $\tau$, only some values of $v$ and $T_{\rm SS}$ are feasible, where the constraints of P1 can be satisfied.
%\end{Definitions}
%mostly depend on $P_T$ and $\tau$,

\subsection{Impact of the Maximum Transmission Power and the Number of Users}\label{results2}

\begin{comment}
\begin{figure*}[t!]
	\centering  
 \setlength\fwidth{0.75\columnwidth}
    \setlength\fheight{0.35\columnwidth}
	\subfigure[$N^*$.] {\input{Figures/Nopt_tau}\label{Nopt_tau}} 
	\subfigure[$P_t^*$.] {\input{Figures/Ptopt_tau}\label{Ptopt_tau}}
 \subfigure[$E_C$.] {\input{Figures/EC_tau}\label{EC_tau}}
 \subfigure[$P_C$.] {\input{Figures/ECT_tau}\label{ECT_tau}}
	\caption{$N^*, P_t^*, E_C$, and $P_C$ as a function of $\tau$ and $K$ at $v=1$ m/s, $T_{\rm SS}=20$ ms, $N_{\rm SS}=8$, and $P_T=30$ dBm.} \label{N_tau}
\end{figure*}
\end{comment}

Fig.~\ref{fig:PT_grouped} depicts $N^*_{\rm gNB}, P_t^*$, $E_C$, and $P_C$ as a function of $P_T$ and $K$. Specifically, we set $\tau=-5$ dB, and consider values of $P_T$ ranging from $10$ to $40$~dBm, while $K\in\{50, 100, 200, 500, 1000\}$. %to provide a comprehensive picture of the impact of these parameters. 
In Fig.~\hyperref[fig:PT_grouped]{\ref*{fig:PT_grouped}(a)} we observe that $N^*_{\rm gNB}$ decreases monotonically when increasing $P_T$, until a minimum value. This is because, by increasing the maximum transmission power (and so the SNR), thus relaxing Constraint $C_2$, we can safely decrease $N^*_{\rm gNB}$ (and so the energy consumption) while still satisfying Constraint $C_1$.
%The rate of decrease of $N^*_{\rm gNB}$ with respect to $P_T$ increases with $K$, since on average we have to satisfy $C_1$ for progressively worse worst cases.

Conversely, Fig.~\hyperref[fig:PT_grouped]{\ref*{fig:PT_grouped}(b)} shows that $P_t^*$ generally increases as $P_T$ increases, until a maximum value after which $P_t^*$ saturates, since higher transmission powers would only entail a higher $E_C$.
  % and the rate of increase is inversely proportional to $P_T$ and tends to zero.
This is explained by the fact that P1 yields the smallest value of $P_t^*$ that meets Constraint $C_1$ for $N^*_{\rm gNB}$. 
For instance, $C_1$ can be met for $N^*_{\rm gNB}=2$ for $P_T\geq 30$~dBm and $K \leq 500$.

Fig.~\hyperref[fig:PT_grouped]{\ref*{fig:PT_grouped}(c)} reports $E_C$ as a function of $K$ and $P_T$.
We see that $E_C$ has a non-monotonic behavior with respect to $P_T$, so there exists an optimal value $P^*_{T,E_C}$ that minimizes the energy consumption. We have that $P^*_{T,E_C}\in\{18, 19, 20, 21, 22\}$~dBm for $K\in\{50, 100, 200, 500, 1000\}$, respectively.
When $P_T<P^*_{T,EC}$, $E_C$ is dominated by $N_{\rm gNB}^*$: since the latter decreases with $P_T$ (see Fig.~\hyperref[fig:PT_grouped]{\ref*{fig:PT_grouped}(a)}), so does $E_C$. 
On the other hand, when $P_T>P^*_{T,E_C}$, $P_t^*$ keeps increasing (see Fig.~\hyperref[fig:PT_grouped]{\ref*{fig:PT_grouped}(b)}) while $N_{\rm gNB}^*$ does not decrease further, so $E_C$ eventually increases. 

Finally, in Fig.~\hyperref[fig:PT_grouped]{\ref*{fig:PT_grouped}(d)} we plot the average power consumption $P_C$ for sending \glspl{ssb} over time. This is defined as the cumulative energy consumed for sending \glspl{ssb} during a complete beam management sweep divided by its duration.  %so it is itself a function of $N_{\rm gNB}$. 
It can be computed as
\begin{align}
	P_C=P_{\rm gNB}T_{\rm SSB}{N_{\rm SS}}/{T_{\rm SS}},
 \label{ECT}
\end{align}
where $P_{\rm gNB}T_{\rm SSB}$ represents the energy consumed for sending one SSB as per Eq.~\eqref{EC1}. 
We observe that $P_C$ has a similar shape as $E_C$ since it also depends on both $N^*_{\rm gNB}$ and $P_t^*$, so there exists an optimal value $P^*_{T,P_C}$ to minimize the power consumption.
%This is because $P_C$ is proportional to both $N^*_{\rm gNB}$ and $P_t^*$, as reported in Eq.~\eqref{ECT}, so 
 We have that $P^*_{T,P_C}\in\{12, 13, 14, 16, 16\}$~dBm for $K\in\{50, 100, 200, 500, 1000\}$, respectively.

\begin{figure}[t!]
	\centering 
     \setlength\fwidth{0.7\columnwidth}
     \setlength\fheight{0.25\columnwidth}
	\input{Figures/Tau_grouped}
	\caption{$N^*_{\rm gNB}, P_t^*, E_C$, and $P_C$ as a function of $\tau$ and $K$, for $v=1$ m/s, $T_{\rm SS}=20$ ms, $N_{\rm SS}=8$, and $P_T=30$~dBm.}
        \label{fig:Tau_grouped}
\end{figure}

As expected, the impact of $K$ is not negligible, and we have that both $E_C$ and $P_C$ increase as $K$ increases. Moreover, as $K$ decreases, the worst-case SNR increases, so $N^*_{\rm gNB}$ and $P_t^*$ have to be higher too to satisfy the constraints in P1. 
Interestingly, when $P_T$ is small, $N^*_{\rm gNB}$ increases quite rapidly with $K$, while the corresponding variation of $P^*_t$ is negligible. This is because $N^*_{\rm gNB}$ has a larger impact on $E_C$ than $P_t^*$. %Consequently, the rate of decreases of $N^*$ increases with $K$. 

%Generally speaking, these trends stem from the fact that, on average, the worst-case SNR decreases as $K$ increases. As a result, higher values of $N^*_{\rm gNB}$ and $P_t^*$ are required to satisfy the constraints in P1. 
%In contrast, for higher values of $P_T$ ($P_T\geq 20$~dBm) %, resulting in higher available transmission power, 
%$N^*_{\rm gNB}$ is already minimized (i.e., $N^*_{\rm gNB}=2$) to minimize $E_C$. Therefore, the increase in $K$ reflects in higher values of $P_t^*$.%., which increase with increasing $K$. This explains why the $E_C$ minimum occurs at larger $P_T$ values as $K$ increases.

% It must be noted that Fig.~\ref{fig:PT_grouped} has been obtained for $v=1$ m/s, $T_{\rm SS}=20$ ms and $N_{\rm SS}=8$.
% While the magnitudes of $N^*_{\rm gNB}$, $P_t^*$, and $E_C$ are independent of the values of these parameters, $P_C$ is a function of the specific values of $T_{\rm SS}$ and $N_{\rm SS}$, as depicted in Eq.~\eqref{ECT}. %, and will change with the change in these parameter values. 
% However, the points of minima for $P_C$ remain the same, since they depend on $N^*_{\rm gNB}$ and $P_t^*$ only. 
% From the above observations, we conclude that both $E_C$ and $P_C$ % in \gls{sma} scenario 
% are minimized for intermediate values of $P_T$. %Thus, we do not require high transmission power at the gNB node to ensure energy efficiency in beam management.

%\begin{Definitions}{}{}
	\Summary
$N^*_{\rm gNB}$ and $P_t^*$ monotonically decrease and increase with $P_T$, respectively. In turn, the shape of both $E_C$ and $P_C$ is non-monotonic, so there exists an optimal (intermediate) value of $P_T$ to optimize the two. Finally, the impact of $K$ is not negligible, and $N^*_{\rm gNB}, P_t^*$, $E_C$ and $P_C$ increase as $K$ increases.
%\end{Definitions}

\subsection{Impact of the SNR Threshold and the Number of Users}
Fig.~\ref{fig:Tau_grouped} reports $N^*_{\rm gNB}, P_t^*$, $E_C$, and $P_C$ as a function of $\tau$ and $K$. 
We set $P_T=30$~dBm, $\tau$ ranging from $0$ to $10$ dB, and $K\in\{50, 100, 200, 500, 1000\}$. 
Figs.~\hyperref[fig:Tau_grouped]{\ref*{fig:Tau_grouped}(a)} and~\hyperref[fig:Tau_grouped]{\ref*{fig:Tau_grouped}(b)} show that both $N^*_{\rm gNB}$ and $P_t^*$ increase monotonically with respect to $\tau$. 
Similarly, Figs.~\hyperref[fig:Tau_grouped]{\ref*{fig:Tau_grouped}(c)} and~\hyperref[fig:Tau_grouped]{\ref*{fig:Tau_grouped}(d)} show that also $E_C$ and $P_C$ increase with $\tau$, and are minimized for $\tau=0$~dB. 
In fact, when $\tau$ is small (so Constraint $C_1$ is light), P1 yields the minimum value of $N^*_{\rm gNB}$ to minimize $E_C$, while $P_t^*$ is large and already close to $P_T$ to meet $C1$. For instance, for $\tau=4$ dB and $K=500$, we have $N^*_{\rm gNB}=2.5$ and $P_t^*\approx 27$ dBm.
%Therefore, the optimal strategy to minimize the $E_C$ is to minimize $N^*_{\rm gNB}$ first, and possibly increase $P_t^*$ to meet Constraint $C_1$. 
%From the results in Fig.~\hyperref[fig:PT_grouped]{\ref*{fig:PT_grouped}}, since in this scenario $P_T=30$ dBm, 
%In fact, we observe that for $\tau=0$ dB, $N^*_{\rm gNB}<3$ for up to $K=1000$ UEs.
When $\tau$ increases, $P_t^*$ keeps increasing.
Still, Constraint $C_2$ upper bounds $P_t^*$ to $P_T$, so $P_t^*$ eventually saturates to $P_T$. In turn, P1 can only be solved by increasing $N^*_{\rm gNB}$, at the expense of $E_C$. 

Notice that the impact of a higher $\tau$ is similar to that of a higher $K$, since both effectively require to either increase the beamforming gain (so increasing $N^*_{\rm gNB}$) or transmit with a higher power (so increasing $P_t^*$) for P1 to be feasible. 
% Move at the end of this subsec?

%Notably, the trend of $E_C$ is similar to that of mainly depends of $N^*_{\rm gNB}$. 
%Since $N^*_{\rm gNB}$ is the dominant term in the expression of $E_C$, the latter increases with $\tau$, and with a trend that resembles that of $N^*_{\rm gNB}$. 
%The rate of increase of $E_C$ with $\tau$ increases further for a larger $K$. This indicates that the energy consumption for a complete beam management process increases with an increase in the SNR threshold and the UE density. 
%On the other hand, $P_C$ exhibits a linear dependence on both $N^*_{\rm gNB}$ and $P_t$. 
%Therefore, the increase in $P_C$ with $\tau$ does not closely match that of $N^*_{\rm gNB}$. 
%These plots have been obtained for a transmission power of $P_T=30$ dBm, such that $K=1000$ UE’s are always feasible up to $\tau=10$ dB. The feasibility limits would vary for smaller $P_T$ values. This has been demonstrated in the following section

%\begin{Definitions}{}{}
\Summary
$N^*_{\rm gNB}$, $P_t^*$, $E_C$, and $P_C$, monotonically increase with $\tau$ since P1 is more constrained. 
The optimal strategy to minimize $E_C$ is to minimize $N^*_{\rm gNB}$ first, and possibly increase $P_t^*$ to meet the SNR constraint.
%\end{Definitions}

%\hl{Temporarily commented 3D plots to speedup compilation}
%\begin{comment}
\begin{figure*}[t!]
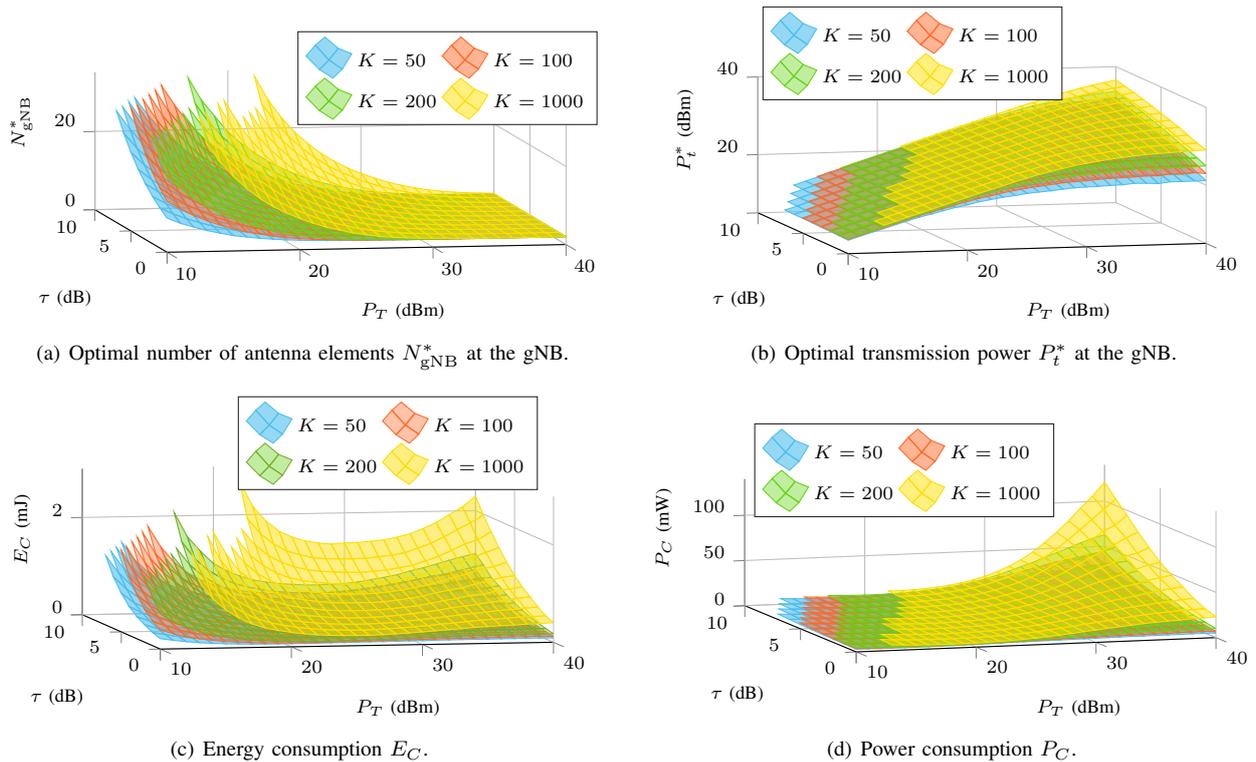

	\centering 
    \setlength\fwidth{0.7\columnwidth}
    \setlength\fheight{0.28\columnwidth}
    \subfigure[Optimal number of antenna elements $N^*_{\rm gNB}$ at the gNB.] {\input{Figures/Nopt_PT_tau}\label{fig:Nopt_PT_tau_3D}} \qquad
    \subfigure[Optimal transmission power $P_t^*$ at the gNB.] {\input{Figures/Ptopt_PT_tau}\label{fig:Ptopt_PT_tau_3D}}
    \subfigure[Energy consumption $E_C$.] {\input{Figures/EC_PT_tau}\label{fig:EC_PT_tau_3D}}\qquad
    \subfigure[Power consumption $P_C$.] {\input{Figures/ECT_PT_tau}\label{fig:ECT_PT_tau_3D}}
	\caption{$N^*_{\rm gNB}, P_t^*, E_C$, and $P_C$ as a function of $P_T$, $\tau$, and $K$. Missing values represent infeasibility.} \label{fig:PT_tau_3D}
\end{figure*}
%\end{comment}

\subsection{Joint Impact of the Transmission Power, the SNR Threshold, and the Number of Users}

Fig.~\ref{fig:PT_tau_3D} depicts the joint impact of $P_T$ and  $\tau$ on $N^*_{\rm gNB}$, $P_t^*$, and $E_C$. 
First, notice that some points are not defined since P1 is infeasible for certain values of $P_T$ and $\tau$.
For instance, for $K=50$ (i.e., with a UE density of $0.0016$/m$^2$), $P_T=10$ dBm and $\tau>7$ dB are not feasible. For $K=1000$, the feasibility domain is even smaller. For instance, $P_T=10$ dBm and $\tau=0$ dB are infeasible. This trend can be explained by the fact that, by increasing $K$, Constraint $C_1$ must be satisfied for an increasing number of realizations. %Nevertheless, the feasibility regions depend on the values of the beam management parameters $T_{\rm SS} and N_{\rm SS}$, and the users' speed $v$.
%The domain of these plots does not necessarily coincide with the full range of values of $P_T$ and $\tau$ since P1 is infeasible for a subset of these input parameters.
%These plots give a comprehensive picture of the impact of different combinations of $P_T$ and $\tau$ on $N^*$, $P_t^*$, $E_C$ and $P_C$. 
%\hl{MP: REVISED UP TO THIS POINT}

Fig.~\ref{fig:Nopt_PT_tau_3D} corroborates that $N^*_{\rm gNB}$ increases with $\tau$ and decreases with $P_T$. 
Notably, the rate of increase of $N^*_{\rm gNB}$ with $\tau$ is lower for higher values of $P_T$. 
In fact, when $P_T$ is large, the system can easily satisfy the SNR constraint regardless of the value of $N^*_{\rm gNB}$, so P1 is tempted to reduce $N^*_{\rm gNB}$ as much as possible to improve $E_C$. For example, $N^*_{\rm gNB}\approx 2$ for $\tau=10$~dB and $P_T=40$ dBm.
%As an illustration, when $P_T=10$ dBm and $\tau=0$ dB, the required number of antenna elements at the gNB is found to be $N^*=8.72$ for a system with $K=50$ UEs. 

Similarly, Fig.~\ref{fig:Ptopt_PT_tau_3D} confirms that $P_t^*$ increases with both $P_T$ and $\tau$. 
In particular, P1 tends to yield the smallest possible value of $P_t^*$ that satisfies $C_1$ to minimize energy consumption, even though the gNB has to use more power when the system is more constrained, i.e., when $\tau$ increases.
Eventually, we observe that $P_t^*$ saturates to $P_T$ when $\tau\geq5$ dB.
%Conversely, $P_t^*$ saturates to a value smaller than $P_T$ for higher values of $\tau$. Indeed, in the latter case $N^*_{\rm gNB}$ is already minimized, and the SNR constraint can be met with $P_t^* < P_T$. Therefore, P1 yields the solution ($N^*_{\rm gNB} = 2$, $P_t^*$), with $P_t^* $ taking the smallest possible value which satisfies the QoS constraints ($C_1$) to minimize energy consumption.
%for a set of $K\in\{50,100, 200, 1000\}$ UEs. 
%For a higher SNR threshold and inadequate transmission power (when the constraints are strict), the value of $P_t^*$ almost approaches the total transmission power $P_T$. The value of $P_t^*$ is considerably less than $P_T$ when the constraints are relaxed.

Finally, Figs.~\ref{fig:EC_PT_tau_3D} and~\ref{fig:ECT_PT_tau_3D} show that both $E_C$ and $P_C$ are non-monotonic with respect to $P_T$, while they both increase as $\tau$ increases.
%Therefore, we conclude that the minimum energy will be consumed at the minimum $\tau$, in this case $\tau=0$ dB. 
%As the value of $K$ increases, there is a corresponding increase in energy consumption at the gNB, since higher $K$ implies higher $N^*_{\rm gNB}$. 
The minimum $E_C$ is observed for $\tau=0$~dB and $P_T\in\{20, 22, 22, 26\}$ dBm for $K\in\{50, 100, 200, 1000\}$, respectively. On a similar note, the minimum $P_C$ is observed for $\tau=0$~dB and $P_T\in\{15, 15, 16, 18\}$ dBm for $K\in\{50, 100, 200, 1000\}$, respectively. Notably, these minimum points are independent of $v$, $T_{\rm SS}$, and $N_{\rm SS}$.

\Summary
$N^*_{\rm gNB}$, $P_t^*$, $E_C$ and $P_C$ increase with $\tau$. 
$N^*_{\rm gNB}$ ($P_t^*$) decreases (increases) with $P_T$, while $E_C$ and $P_C$ show a non-monotonic trend with respect to $P_T$.

\subsection{Feasibility Regions} %in terms of $N_{\rm SS}$ and $T_{\rm SS}$, and as a function of $K$}
\label{sec_FL}
Based on the above discussion, we infer that, for each combination of $P_T$ and $\tau$, there is a limit on the values of $T_{\rm SS}$, $N_{\rm SS}$, and $v$ for which the problem P1~\eqref{OP1} is feasible, that is when the SNR constraint is met. 
%which prevents the misalignment of the UEs with respect to the corresponding associated beam also increases.
As described in Sec.~\ref{feasibility_limit}, 
%both $v$ and $T_{\rm SS}$ exert an equal impact on the angular offset. Furthermore, 
these values identify the so-called feasibility regions of P1, that is the set of feasible beam management parameters $\{T_{\rm SS}$, $N_{\rm SS}\}$ for RedCap, and the corresponding maximum speed $v$ and number of UEs $K$ that can be tolerated.
We recall that, as observed in Sec.~\ref{modeling}, the angular offset is defined in terms of the product of $v$ and $T_{\rm SS}$, so we characterize the shape of the feasibility regions also based on this product. % of $v$ and $T_{\rm SS}$. 
%For example, from Fig.~\ref{Nopt_Tss}, for $P_T=18$ dBm, $\tau=7$ dB, and $N_{\rm SS}=8$, $v>2$ m/s is infeasible at $T_{\rm SS}=160$ ms. Similarly, $v>4$ m/s is infeasible at $T_{\rm SS}=80$ ms. Therefore, we conclude that the feasibility region is upper bounded by $vT_{\rm SS}\leq 0.32$ m. 
%The limits for various values of $N_{\rm SS}$ and $K$ can be derived using a similar methodology. 
\begin{table}[h]
	\centering
	\renewcommand{\arraystretch}{1.1}
	\caption{Feasibility regions for $P_T=18$~dBm and $\tau=7$~dB.}
	\begin{tabular}{ c|c|c|c|c|c }
		\hline\
		& \multicolumn{5}{c}{Upper bound on $vT_{\rm SS}$} \\
		\hline
		\diagbox{$N_{\rm SS}$}{$K$} & $50$ &$100$ & $200$ & $500$ & $1000$\\
		\hline
		$8$   &$0.32$ m  &$0.16$ m  & $0.08$ m & $0.03$ m & NF\\
		\hline
		$16$   &$0.64$ m &$0.32$ m  & $0.16$ m & $0.06$ m & NF\\
		\hline
		$32$   &$1.68$ m & $0.72$ m & $0.32$ m & $0.12$ m & NF\\
		\hline
		$64$   &$4.80$ m & $1.76$ m & $0.64$ m & $0.32$ m & NF\\
		\hline
	\end{tabular}
	\label{FR_K_table}
 %\vspace{-0.33cm}
\end{table}

\begin{figure}
	\centering  
 \setlength\fwidth{0.75\columnwidth}
    \setlength\fheight{0.3\columnwidth}
	% This file was created by matlab2tikz.
%
%The latest updates can be retrieved from
%  http://www.mathworks.com/matlabcentral/fileexchange/22022-matlab2tikz-matlab2tikz
%where you can also make suggestions and rate matlab2tikz.
%
\definecolor{mycolor1}{rgb}{0.00000,0.44700,0.74100}%
\definecolor{mycolor2}{rgb}{0.85000,0.32500,0.09800}%
\definecolor{mycolor3}{rgb}{0.92900,0.69400,0.12500}%
\definecolor{mycolor4}{rgb}{0.46667,0.67451,0.18824}%
\begin{tikzpicture}

\begin{axis}[%
width=\fwidth,
height=\fheight,
at={(0\fwidth,0\fheight)},
scale only axis,
area style,
%stack plots=y,
xmin=8,
xmax=64,
xtick={8,16,32,64},
xlabel={$N_{\rm SS}$},
ymin=0,
ymax=160,
ytick={5,20,40,80,160},
yticklabels={{5},{10},{20},{40},{80},{160}},
ylabel={$T_{\rm SS}$ (ms)},
%axis background/.style={fill=white},
xmajorgrids,
ymajorgrids,
legend style={at={(0.5,0.8)}, anchor=south, legend cell align=left, align=left, draw=white!15!black, legend columns = 2}
]
\addplot[fill=mycolor1, draw=black] table[row sep=crcr]{%
8	    40\\
15.99   40\\
16	    80\\
31.99	80\\
32	    160\\
63.99	160\\
64	    160\\
}
\closedcycle;
\addlegendentry{$K=50$}

\addplot[preaction={fill, mycolor2}, pattern={north east lines}, draw=black] table[row sep=crcr]{%
8	    20\\
15.99	20\\
16	    40\\
31.99	40\\
32	    80\\
63.99	80\\
64	    160\\
}
\closedcycle;
\addlegendentry{$K=100$}

\addplot[preaction={fill, mycolor3}, pattern={dots}, draw=black] table[row sep=crcr]{%
8	10\\
15.99	10\\
16	20\\
31.99	20\\
32	40\\
63.99	40\\
64	160\\
}
\closedcycle;
\addlegendentry{$K=200$}

\addplot[preaction={fill, mycolor4}, pattern={grid}, draw=black] table[row sep=crcr]{%
8	5\\
15.99	5\\
16	10\\
31.99	10\\
32	20\\
63.99	20\\
64	40\\
}
\closedcycle;
\addlegendentry{$K=500$}

\end{axis}

%\begin{axis}[%
%width=1.227\fwidth,
%height=1.227\fheight,
%at={(-0.16\fwidth,-0.135\fheight)},
%scale only axis,
%xmin=0,
%xmax=1,
%ymin=0,
%ymax=1,
%axis line style={draw=none},
%ticks=none,
%axis x line*=bottom,
%axis y line*=left
%]
%\node[below right, align=left, inner sep=0]
%at (rel axis cs:0.467,0.777) {$N^*=7.4$, $P_t^*=16.4$ dBm, $\rm EC=226.7\mu$J};
%\node[below right, align=left, inner sep=0]
%at (rel axis cs:0.474,0.242) {$N^*=16.2$, $P_t^*=16.8$ dBm, $\rm EC=691.2\mu$J};
%\node[below right, align=left, inner sep=0]
%at (rel axis cs:0.466,0.395) {$N^*=11$, $P_t^*=16.6$ dBm, $\rm EC=385.2\mu$J};
%\node[below right, align=left, inner sep=0]
%at (rel axis cs:0.475,0.157) {$N^*=27$, $P_t^*=17$ dBm, $\rm EC=1527.5\mu$J};
%\end{axis}
\end{tikzpicture}%
	\caption{Feasibility region for $P_T=18$ dBm, $\tau=7$ dB, $v=5$ m/s, for $K\in\{50, 100, 200, 500\}$.}\label{fig:FR_K}
\end{figure}
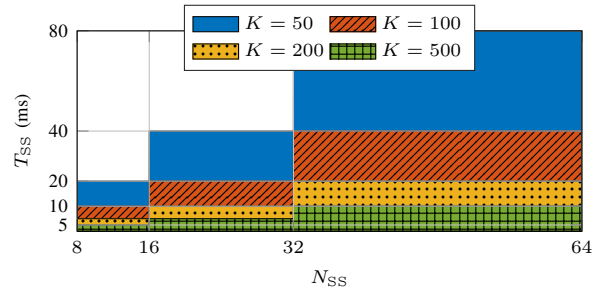

Specifically, Fig.~\ref{fig:FR_K} and Table~\ref{FR_K_table} depict and report, respectively, the feasibility regions for %a \gls{sma} scenario with 
$P_T = 18$~dBm, $\tau=7$~dB, $K\in\{50, 100, 200, 500, 1000\}$, %in terms of the highest product of $v$ and $T_{\rm SS}$ supported by the system, 
and considering $v=5$~m/s. 
%\textbf{Impact of $K$ on the feasibility limits}:
We observe that the feasibility region becomes smaller, i.e., more values of $v$ and $T_{\rm SS}$ become infeasible, %to smaller $T_{\rm SS}$ values 
as $K$ increases. This is because increasing the number of UEs translates into a tighter constraint on the SNR ($C_1$), so P1 tends to increase $N^*_{\rm gNB}$ to increase the beamforming gain. However, this also yields a narrower beam, thereby reducing the speed and the burst periodicity which can be sustained to avoid the misalignment of the UEs with respect to the corresponding associated beam.

 %The feasibility regions are derived by considering the maximum velocity, $v=30$ m/s and burst periodicity, $T_{\rm SS}=160$ ms. Fig.~\ref{FR_K} gives a plot of the feasibility region for $K=\{50, 100, 200, 500\}$ at UE mobility of $v=5$ m/s. It was generated using the values in Table~\ref{FR_K_table}, and are intended to provide guidelines towards the optimal 5G NR beam management configurations to minimize the energy consumption for a \gls{sma} scenario.

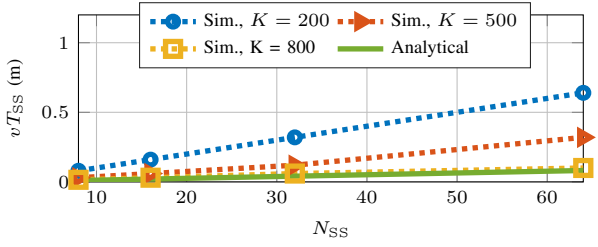
\begin{figure}
	\centering 
    \setlength\fwidth{0.75\columnwidth}
    \setlength\fheight{0.25\columnwidth}
    \begin{tikzpicture}

\definecolor{mycolor1}{rgb}{0.00000,0.44700,0.74100}%
\definecolor{mycolor2}{rgb}{0.85000,0.32500,0.09800}%
\definecolor{mycolor3}{rgb}{0.92900,0.69400,0.12500}%
\definecolor{mycolor4}{rgb}{0.46667,0.67451,0.18824}%

\begin{axis}[%
width=\fwidth,
height=\fheight,
at={(0\fwidth,0\fheight)},
scale only axis,
xmin=8,
xmax=64,
xlabel={$N_{\rm SS}$},
ymin=0,
ymax=1.2,
ylabel={$vT_{\rm SS}$ (m)},
axis background/.style={fill=white},
xmajorgrids,
ymajorgrids,
legend style={at={(0.12,0.68)}, anchor=south west, legend cell align=left, align=left, draw=white!15!black, legend columns=2}
]
%\addplot [color=blue, dotted, line width=2.0pt, mark=diamond, mark %options={solid, blue}]
%  table[row sep=crcr]{%
%8	0.16\\
%16	0.32\\
%32	0.72\\
%64	1.76\\
%};
%\addlegendentry{S, $K=100$}

\addplot [color=mycolor1, dotted, line width=2.0pt, mark=o, mark options={solid, mycolor1}]
  table[row sep=crcr]{%
8	0.08\\
16	0.16\\
32	0.32\\
64	0.64\\
};
\addlegendentry{Sim., $K=200$}

\addplot [color=mycolor2, dotted, line width=2.0pt, mark size=2.7pt, mark=triangle, mark options={solid, rotate=270, mycolor2}]
  table[row sep=crcr]{%
8	0.03\\
16	0.06\\
32	0.12\\
64	0.32\\
};
\addlegendentry{Sim., $K=500$}

\addplot [color=mycolor3, dotted, line width=2.0pt, mark size=2.8pt, mark=square, mark options={solid, mycolor3}]
  table[row sep=crcr]{%
8	0.015\\
16	0.03\\
32	0.06\\
64	0.1\\
};
\addlegendentry{Sim., K = 800}

\addplot [color=mycolor4, line width=2.0pt]
  table[row sep=crcr]{%
8	0.0099180480046013\\
16	0.0203917777097806\\
32	0.0410931449570502\\
64	0.0818192064971695\\
};
\addlegendentry{Analytical}

\end{axis}
\end{tikzpicture}%
	\caption{Feasibility regions for $P_T=18$~dBm, $\tau=7$~dB, and $v=5$~m/s. Dashed lines are relative to the numerical results obtained via simulations for $K\in\{50, 100, 200, 500\}$, while the straight line represents the analytical feasibility lower bound for $K \to +\infty$ based on Eq.~\eqref{vTss_lim_det}.}\label{FL_K}
\end{figure}

Fig.~\ref{FL_K} compares the feasibility limits %on $vT_{\rm SS}$ as 
determined through numerical evaluation (and reported in Table~\ref{FR_K_table}), and by using %the analytical expression derived in Sec.~\ref{feasibility_limit}, 
Eq.~\eqref{vTss_lim_det}, for $K\in\{200, 500, 800\}$. 
We can see that Eq.~\eqref{vTss_lim_det}, which effectively models the feasibility region for $K \to \infty$, provides a worst-case lower bound of the values obtained via simulation. As such, this bound is observed to be tight for $K=800$, while instead there is a gap between the analytical model and the simulations as $K$ decreases. This is due to the fact that, when $K$ is small, the probability that a UE is located at the cell edge, and so measures an SNR which is close to that of the worst case, is small, so the corresponding feasibility regions are less constrained than those of the lower bound.
%We observe that the analytical bound is close at small values of $N_{\rm SS}$ and loosens as $N_{\rm SS}$ increases. We also observe a difference in the feasibility limits for $K=100$ and $K=200$ where the simulation indicates that the problem is feasible at $N_{\rm SS}=64$, whereas the analytical problem indicates otherwise. The difference arises due to the inclusion of cases in \gls{mc} simulation where smaller values of $N_{\rm gNB}$ are feasible. Therefore, after obtaining the optimal antenna, the derived expression in Eq.~\eqref{vTss_lim} can be used to determine the limit on mobility and burst periodicity.

Finally, we observe that the size of the feasibility regions is inversely proportional to $v$ and $T_{\rm SS}$. Indeed, if the UEs move faster (i.e., $v$ increases), or if the beam management process takes longer (i.e., $T_{\rm SS}$ increases), the probability that the UEs would lose beam alignment also increases. As such, the set of possible values of $v$ and $T_{\rm SS}$ to make the problem feasible is smaller. Still, the impact of $T_{\rm SS}$ on the feasibility regions is zero for $N_{\rm SS} \geq S_D$, i.e., if the transmission of the SSBs requires exactly one burst, as we can see from Eq.~\eqref{T_BM}.

%\begin{Definitions}{}{}
\Summary
The size of the feasibility regions, i.e., the set of feasible values of RedCap beam managemenet parameters $\{T_{\rm SS}$, $N_{\rm SS}\}$, depends on $v$ and $K$.
In any case, decreasing the beam management period,
%as stated in Eq.~\eqref{OP2},
i.e., decreasing $T_{\rm SS}$ and maximizing $N_{\rm SS}$, promotes feasibility. 

%which prevents the misalignment of the UEs with respect to the corresponding associated beam also increases.
%\end{Definitions}

\begin{figure}[h!]
	\centering  
    \setlength\fwidth{0.75\columnwidth}
    \setlength\fheight{0.3\columnwidth}
	\subfigure[Optimal number of antenna elements $N^*_{\rm gNB}$ at the gNB.] {% This file was created by matlab2tikz.
%
%The latest updates can be retrieved from
%  http://www.mathworks.com/matlabcentral/fileexchange/22022-matlab2tikz-matlab2tikz
%where you can also make suggestions and rate matlab2tikz.
%
\definecolor{mycolor1}{rgb}{0.46667,0.67451,0.18824}%
\definecolor{mycolor2}{rgb}{0.92941,0.69412,0.12549}%
\begin{tikzpicture}

\begin{axis}[%
width=\fwidth,
height=\fheight,
at={(0\fwidth,0\fheight)},
scale only axis,
xmin=10,
xmax=40,
xlabel={$P_T$ (dBm)},
ymin=0,
ymax=40,
ylabel={$N^*_{\rm gNB}$},
axis background/.style={fill=white},
xmajorgrids,
ymajorgrids,
legend style={legend cell align=left, align=left, draw=white!15!black}
]
\addplot [color=blue, line width=1.5pt, mark=triangle, mark repeat=3, mark options={solid, rotate=270, blue}]
  table[row sep=crcr]{%
15	27.71758\\
16	23.08608\\
17	19.3593\\
18	16.21382\\
19	13.57652\\
20	11.5119\\
21	9.70004\\
22	8.22324\\
23	6.94136\\
24	5.87074\\
25	4.971\\
26	4.24512\\
27	3.6609\\
28	3.20174\\
29	2.86464\\
30	2.6047\\
31	2.4298\\
32	2.29892\\
33	2.20912\\
34	2.14776\\
35	2.10298\\
36	2.07112\\
37	2.04568\\
38	2.0317\\
39	2.02148\\
40	2.01502\\
};
\addlegendentry{$P_{\rm MD}=0$}

\addplot [color=red, line width=1.5pt, mark size=3.0pt, mark repeat=3, mark=asterisk, mark options={solid, red}]
  table[row sep=crcr]{%
11	36.1522\\
12	29.5844\\
13	24.421\\
14	20.173\\
15	16.7194\\
16	13.8004\\
17	11.501\\
18	9.5342\\
19	8.0456\\
20	6.6846\\
21	5.483\\
22	4.6096\\
23	3.8612\\
24	3.2488\\
25	2.7892\\
26	2.4302\\
27	2.246\\
28	2.1166\\
29	2.058\\
30	2.0266\\
31	2.0108\\
32	2.0042\\
33	2.0012\\
34	2.0004\\
35	2\\
36	2\\
37	2\\
38	2\\
39	2\\
40	2\\
};
\addlegendentry{$P_{\rm MD}=0.01$}

\addplot [color=mycolor1, line width=1.5pt, mark=square, mark repeat=3, mark options={solid, mycolor1}]
  table[row sep=crcr]{%
10	20.0186\\
11	16.3956\\
12	13.3818\\
13	10.9788\\
14	9.1066\\
15	7.5504\\
16	6.1726\\
17	5.0116\\
18	4.115\\
19	3.4114\\
20	2.804\\
21	2.3606\\
22	2.107\\
23	2.024\\
24	2.0028\\
25	2.0002\\
26	2\\
27	2\\
28	2\\
29	2\\
30	2\\
31	2\\
32	2\\
33	2\\
34	2\\
35	2\\
36	2\\
37	2\\
38	2\\
39	2\\
40	2\\
};
\addlegendentry{$P_{\rm MD}=0.05$}

\addplot [color=mycolor2, line width=1.5pt, mark=diamond, mark repeat=3, mark options={solid, mycolor2}]
  table[row sep=crcr]{%
10	12.0818\\
11	9.8926\\
12	8.1862\\
13	6.7562\\
14	5.4298\\
15	4.4296\\
16	3.6398\\
17	2.9972\\
18	2.4366\\
19	2.1086\\
20	2.014\\
21	2.0008\\
22	2\\
23	2\\
24	2\\
25	2\\
26	2\\
27	2\\
28	2\\
29	2\\
30	2\\
31	2\\
32	2\\
33	2\\
34	2\\
35	2\\
36	2\\
37	2\\
38	2\\
39	2\\
40	2\\
};
\addlegendentry{$P_{\rm MD}=0.1$}

\end{axis}
\end{tikzpicture}%
\label{Nopt_pmd}} 
	\subfigure[Optimal transmission power $P_t^*$ at the gNB.] {% This file was created by matlab2tikz.
%
%The latest updates can be retrieved from
%  http://www.mathworks.com/matlabcentral/fileexchange/22022-matlab2tikz-matlab2tikz
%where you can also make suggestions and rate matlab2tikz.
%
\definecolor{mycolor1}{rgb}{0.46667,0.67451,0.18824}%
\definecolor{mycolor2}{rgb}{0.92941,0.69412,0.12549}%
\begin{tikzpicture}

\begin{axis}[%
width=\fwidth,
height=\fheight,
at={(0\fwidth,0\fheight)},
scale only axis,
xmin=10,
xmax=40,
xlabel={$P_T$ (dBm)},
ymin=5,
ymax=35,
ylabel={$P_t^*$  (dBm)},
axis background/.style={fill=white},
xmajorgrids,
ymajorgrids,
legend style={at={(0.02,0.53)}, anchor=south west, legend cell align=left, align=left, draw=white!15!black}
]
\addplot [color=blue, line width=1.5pt, mark=triangle, mark repeat=3, mark options={solid, rotate=270, blue}]
  table[row sep=crcr]{%
15	13.9766891762552\\
16	14.9340407003586\\
17	15.8771271441236\\
18	16.8209894867542\\
19	17.7660068004045\\
20	18.6935829239417\\
21	19.6339117910395\\
22	20.5623529595959\\
23	21.4993408240182\\
24	22.399826158402\\
25	23.3186731226942\\
26	24.195082197616\\
27	25.0688264163338\\
28	25.8894481535278\\
29	26.6665132938564\\
30	27.3963919096893\\
31	28.0555342111134\\
32	28.6092106025549\\
33	29.1463778992346\\
34	29.604666208923\\
35	30.0170567180483\\
36	30.3574056020646\\
37	30.6567107026102\\
38	30.8872987558541\\
39	31.1267496689392\\
40	31.2897642152332\\
};
\addlegendentry{$P_{\rm MD}=0$}

\addplot [color=red, line width=1.5pt, mark size=3.0pt, mark=asterisk, mark repeat=3, mark options={solid, red}]
  table[row sep=crcr]{%
11	10.2720595793249\\
12	11.2338832440378\\
13	12.2032563323507\\
14	13.1473861266224\\
15	14.117540947783\\
16	15.0597921623183\\
17	16.024891256015\\
18	16.9594798897464\\
19	17.8858177908476\\
20	18.8320162995384\\
21	19.7957314116244\\
22	20.6937544384584\\
23	21.6050423777259\\
24	22.4572916085367\\
25	23.3065221689308\\
26	24.1068093988058\\
27	24.7338829116327\\
28	25.2329591849664\\
29	25.5463829331861\\
30	25.8156331211337\\
31	25.9344042936364\\
32	25.9725836161984\\
33	26.0181111396322\\
34	26.0365001030094\\
35	26.1088118899181\\
36	26.0546507086743\\
37	26.0640957475697\\
38	26.0632387797053\\
39	26.0935804211912\\
40	25.9907143275344\\
};
\addlegendentry{$P_{\rm MD}=0.01$}

\addplot [color=mycolor1, line width=1.5pt, mark=square, mark repeat=3, mark options={solid, mycolor1}]
  table[row sep=crcr]{%
10	9.38180281612993\\
11	10.3587028137278\\
12	11.3233289319688\\
13	12.2700266070007\\
14	13.2234989281103\\
15	14.1544600222003\\
16	15.1366364262058\\
17	16.0764101371461\\
18	16.9854101102981\\
19	17.876467272989\\
20	18.7806040777707\\
21	19.6471657249375\\
22	20.2759775494445\\
23	20.6254554948747\\
24	20.7119618686546\\
25	20.7376480805621\\
26	20.752595185099\\
27	20.7478935360179\\
28	20.7415683019418\\
29	20.7416937622649\\
30	20.7200684139707\\
31	20.703644508996\\
32	20.7491367746625\\
33	20.7404626133679\\
34	20.7157703389108\\
35	20.7107680702544\\
36	20.7371311795718\\
37	20.733364452252\\
38	20.7349478570541\\
39	20.7243390778916\\
40	20.7199247002722\\
};
\addlegendentry{$P_{\rm MD}=0.05$}

\addplot [color=mycolor2, line width=1.5pt, mark=diamond, mark repeat=3, mark options={solid, mycolor2}]
  table[row sep=crcr]{%
10	9.39027518857346\\
11	10.354852556233\\
12	11.2904723473702\\
13	12.2380480027303\\
14	13.2186124686749\\
15	14.1352503230721\\
16	15.0471240367846\\
17	15.9420765269485\\
18	16.8582941706683\\
19	17.5921652910777\\
20	17.8861519752095\\
21	17.9321546634063\\
22	17.9409961490343\\
23	17.9556520741035\\
24	17.9660781359917\\
25	17.9557174679918\\
26	17.9639355577983\\
27	17.9832687672502\\
28	17.9767307905238\\
29	17.9588535968868\\
30	17.9522861023039\\
31	17.9425671559115\\
32	17.9640141030884\\
33	17.9487512397168\\
34	17.9461511575258\\
35	17.9529126495169\\
36	17.9443311027905\\
37	17.9342859517903\\
38	17.9568907033063\\
39	17.9515841056317\\
40	17.9452929547545\\
};
\addlegendentry{$P_{\rm MD}=0.1$}

\end{axis}
\end{tikzpicture}%
\label{Ptopt_pmd}}
    \subfigure[Energy consumption $E_C$.] {% This file was created by matlab2tikz.
%
%The latest updates can be retrieved from
%  http://www.mathworks.com/matlabcentral/fileexchange/22022-matlab2tikz-matlab2tikz
%where you can also make suggestions and rate matlab2tikz.
%
\definecolor{mycolor1}{rgb}{0.46667,0.67451,0.18824}%
\definecolor{mycolor2}{rgb}{0.92941,0.69412,0.12549}%
\begin{tikzpicture}

\begin{axis}[%
width=0.975\fwidth,
height=\fheight,
at={(0\fwidth,0\fheight)},
scale only axis,
xmin=10,
xmax=40,
xlabel={$P_T$ (dBm)},
ymin=0,
ymax=2.5,
ylabel={$E_C$ (mJ)},
axis background/.style={fill=white},
xmajorgrids,
ymajorgrids,
legend style={legend cell align=left, align=left, draw=white!15!black}
]
\addplot [color=blue, line width=1.5pt, mark=triangle, mark repeat=3, mark options={solid, rotate=270, blue}]
  table[row sep=crcr]{%
15	1.44458955292716\\
16	1.09874724863417\\
17	0.860888483288736\\
18	0.68853758849925\\
19	0.564372283571785\\
20	0.482676605878053\\
21	0.420557220907296\\
22	0.378334764277501\\
23	0.346792020271245\\
24	0.324171891171329\\
25	0.311055397653509\\
26	0.304992268825861\\
27	0.30686382266256\\
28	0.314480537883853\\
29	0.330111451656158\\
30	0.351380021881942\\
31	0.379176960830137\\
32	0.406509334484374\\
33	0.441798884859047\\
34	0.476611619994606\\
35	0.512157672932031\\
36	0.545578907212874\\
37	0.57715828975683\\
38	0.604306658779511\\
39	0.634463161383415\\
40	0.656301491782198\\
};
\addlegendentry{$P_{\rm MD}=0$}

\addplot [color=red, line width=1.5pt, mark size=3.0pt, mark repeat=3, mark=asterisk, mark options={solid, red}]
  table[row sep=crcr]{%
11	2.13318872360015\\
12	1.51956023376873\\
13	1.11463585273252\\
14	0.831461549830269\\
15	0.635753978875518\\
16	0.493179254988809\\
17	0.397214822687859\\
18	0.325886769825018\\
19	0.280155642332405\\
20	0.241477512045518\\
21	0.211450847931503\\
22	0.19433787064294\\
23	0.182888090104935\\
24	0.174754580431681\\
25	0.173477875295427\\
26	0.175774832838997\\
27	0.184303104703598\\
28	0.192822328034688\\
29	0.200495617467386\\
30	0.209035248173943\\
31	0.212810265612267\\
32	0.213959516618363\\
33	0.215672085842105\\
34	0.216393900757767\\
35	0.219490838957018\\
36	0.217151204425657\\
37	0.217557110088067\\
38	0.217520245005216\\
39	0.218829924012879\\
40	0.214426595496504\\
};
\addlegendentry{$P_{\rm MD}=0.01$}

\addplot [color=mycolor1, line width=1.5pt, mark=square, mark repeat=3, mark options={solid, mycolor1}]
  table[row sep=crcr]{%
10	0.764002227745953\\
11	0.560459324870712\\
12	0.415604850847922\\
13	0.316435843470885\\
14	0.250805024384272\\
15	0.201099550157407\\
16	0.161277473939021\\
17	0.13254234929014\\
18	0.113649524592702\\
19	0.100704296144761\\
20	0.0905683831434272\\
21	0.0850522184074276\\
22	0.0834851389684479\\
23	0.0849083898676014\\
24	0.0852977743236694\\
25	0.0855275610370194\\
26	0.0857092569750026\\
27	0.085649751268027\\
28	0.0855697984670539\\
29	0.0855713831903988\\
30	0.0852989025702615\\
31	0.0850928650666463\\
32	0.0856654798644435\\
33	0.0855558341944194\\
34	0.0852449080232977\\
35	0.0851821342972446\\
36	0.0855137814021236\\
37	0.0854662727260747\\
38	0.0854862387472001\\
39	0.0853526057176009\\
40	0.0852970963039889\\
};
\addlegendentry{$P_{\rm MD}=0.05$}

\addplot [color=mycolor2, line width=1.5pt, mark=diamond, mark repeat=3, mark options={solid, mycolor2}]
  table[row sep=crcr]{%
10	0.342229769297555\\
11	0.259833381598736\\
12	0.203050356030811\\
13	0.158851542954874\\
14	0.123423858205708\\
15	0.100925860508792\\
16	0.0851554518280652\\
17	0.073563531032185\\
18	0.0642144863564733\\
19	0.0600468931805545\\
20	0.0594940859480251\\
21	0.0594608362418884\\
22	0.0594987866540831\\
23	0.059596090454702\\
24	0.0596655114544899\\
25	0.0595965253548835\\
26	0.0596512316809454\\
27	0.0597803382920486\\
28	0.0597366135415656\\
29	0.0596173897797252\\
30	0.0595737139966558\\
31	0.0595092011967723\\
32	0.0596517550422169\\
33	0.0595502334412394\\
34	0.0595329744203789\\
35	0.0595778778717292\\
36	0.0595208992708951\\
37	0.0594543456988569\\
38	0.059604329035514\\
39	0.0595690494169876\\
40	0.059527280045305\\
};
\addlegendentry{$P_{\rm MD}=0.1$}

\end{axis}

\begin{axis}[%
width=1.258\fwidth,
height=1.258\fheight,
at={(-0.164\fwidth,-0.163\fheight)},
scale only axis,
xmin=0,
xmax=1,
ymin=0,
ymax=1,
axis line style={draw=none},
ticks=none,
axis x line*=bottom,
axis y line*=left
]
\end{axis}
\end{tikzpicture}%
\label{EC_pmd}}
    \subfigure[Power consumption $P_C$.] {% This file was created by matlab2tikz.
%
%The latest updates can be retrieved from
%  http://www.mathworks.com/matlabcentral/fileexchange/22022-matlab2tikz-matlab2tikz
%where you can also make suggestions and rate matlab2tikz.
%
\definecolor{mycolor1}{rgb}{0.46667,0.67451,0.18824}%
\definecolor{mycolor2}{rgb}{0.92941,0.69412,0.12549}%
\begin{tikzpicture}

\begin{axis}[%
width=\fwidth,
height=\fheight,
at={(0\fwidth,0\fheight)},
scale only axis,
xmin=10,
xmax=40,
xlabel={$P_T$ (dBm)},
ymin=0,
ymax=40,
ylabel={$P_C$ (mW)},
axis background/.style={fill=white},
xmajorgrids,
ymajorgrids,
legend style={at={(0.02,0.38)}, anchor=south west, legend cell align=left, align=left, draw=white!15!black}
]
\addplot [color=blue, line width=1.5pt, mark=triangle, mark repeat=3, mark options={solid, rotate=270, blue}]
  table[row sep=crcr]{%
15	6.37879375279395\\
16	5.82783488799591\\
17	5.45310063164025\\
18	5.21659642142162\\
19	5.11909896458381\\
20	5.1770918067567\\
21	5.37073242237241\\
22	5.72208280808163\\
23	6.24880467114976\\
24	6.94372604407584\\
25	7.88705299185989\\
26	9.04393661491434\\
27	10.5011395290798\\
28	12.1962042050813\\
29	14.1537172628882\\
30	16.35832779317\\
31	18.7095168912468\\
32	20.9856704971757\\
33	23.4977049408304\\
34	25.9044789328308\\
35	28.3008547882357\\
36	30.4586854800589\\
37	32.501924730773\\
38	34.1760024279325\\
39	36.0118955498068\\
40	37.3210572893749\\
};
\addlegendentry{$P_{\rm MD}=0$}

\addplot [color=red, line width=1.5pt, mark size=3.0pt, mark repeat=3, mark=asterisk, mark options={solid, red}]
  table[row sep=crcr]{%
11	7.29981079221221\\
12	6.35735475341135\\
13	5.6493922985635\\
14	5.10128480062523\\
15	4.70556224583503\\
16	4.42094355597936\\
17	4.27740544675509\\
18	4.22834210706755\\
19	4.31088809521819\\
20	4.49648324721332\\
21	4.8130304320925\\
22	5.25756224014867\\
23	5.86605387988437\\
24	6.60080937721433\\
25	7.53494953087846\\
26	8.62453533661753\\
27	9.65290669045621\\
28	10.5901446434887\\
29	11.2409542233487\\
30	11.8422540403467\\
31	12.1193772815574\\
32	12.2098609208038\\
33	12.3195668745978\\
34	12.3642699997856\\
35	12.5423336546867\\
36	12.4086402528947\\
37	12.4318348621753\\
38	12.4297282860123\\
39	12.5045670864502\\
40	12.2529483140859\\
};
\addlegendentry{$P_{\rm MD}=0.01$}

\addplot [color=mycolor1, line width=1.5pt, mark=square, mark repeat=3, mark options={solid, mycolor1}]
  table[row sep=crcr]{%
10	4.76121968411932\\
11	4.26067188015527\\
12	3.86758074965252\\
13	3.58453402449681\\
14	3.40541941425002\\
15	3.29829498853632\\
16	3.26050215046536\\
17	3.28991148790304\\
18	3.40101441268796\\
19	3.59319074959371\\
20	3.87465525927465\\
21	4.24703140329137\\
22	4.58752735218595\\
23	4.81070447810735\\
24	4.86883774003462\\
25	4.88690534578776\\
26	4.897671827143\\
27	4.89427150103011\\
28	4.88970276954593\\
29	4.88979332516564\\
30	4.87422300401494\\
31	4.86244943237979\\
32	4.8951702779682\\
33	4.88890481110968\\
34	4.8711376013313\\
35	4.86755053127112\\
36	4.88650179440707\\
37	4.88378701291855\\
38	4.88492792841143\\
39	4.87729175529148\\
40	4.87411978879936\\
};
\addlegendentry{$P_{\rm MD}=0.05$}

\addplot [color=mycolor2, line width=1.5pt, mark=diamond, mark repeat=3, mark options={solid, mycolor2}]
  table[row sep=crcr]{%
10	3.53847937907559\\
11	3.25820119286232\\
12	3.06415444339776\\
13	2.93049408625051\\
14	2.83814147378856\\
15	2.81429412127247\\
16	2.85256475897098\\
17	2.94684860790521\\
18	3.10427052925816\\
19	3.28971681509293\\
20	3.38144700673856\\
21	3.39670870733378\\
22	3.39993066594761\\
23	3.40549088312583\\
24	3.40945779739942\\
25	3.40551573456478\\
26	3.40864181033974\\
27	3.41601933097421\\
28	3.41352077380375\\
29	3.40670798741287\\
30	3.40421222838033\\
31	3.4005257826727\\
32	3.40867171669811\\
33	3.40287048235654\\
34	3.40188425259308\\
35	3.40445016409882\\
36	3.40119424405115\\
37	3.39739118279182\\
38	3.40596165917223\\
39	3.40394568097072\\
40	3.40155885973171\\
};
\addlegendentry{$P_{\rm MD}=0.1$}
\end{axis}

\end{tikzpicture}%
\label{ECt_pmd}}
	\caption{$N^*_{\rm gNB}$, $P_t^*$, $E_C$, and $P_C$ as a function of $P_T$ and $P_{\rm MD}$ for $v=1$ m/s, $T_{\rm SS}=20$ ms, $N_{\rm SS}=8$, $\tau=7$ dB, and $K=200$. Missing values represent infeasibility.}\label{NPt_pmd}
\end{figure}
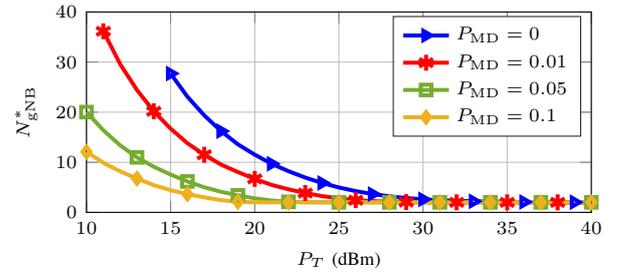
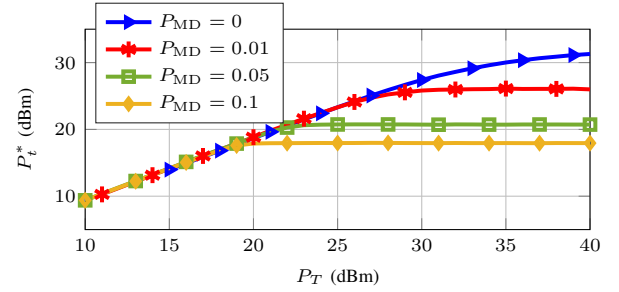
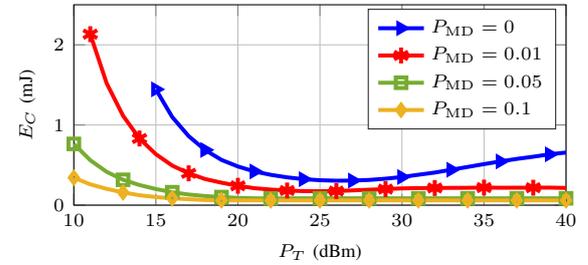
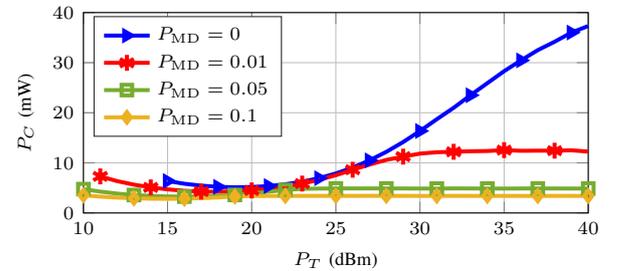

\subsection{Arbitrary Misdetection Probability}
In this section we extend the previous analysis to P2, therefore exploring the impact of a non-zero misdetection probability ($P_{\rm MD}>0$) on the optimal gNB configuration.
In particular, Fig.~\ref{NPt_pmd} reports $N^*_{\rm gNB}$ and $P_t^*$ as a function of $P_T$, for $P_{\rm MD}$ ranging from $0$ to $0.1$, $v=1$~m/s, $T_{\rm SS}=20$~ms, $N_{\rm SS}=8$, $\tau=7$~dB, and $K=200$. 
Overall, it can be observed that $N^*_{\rm gNB}$ and $P_t^*$ decrease as $P_{\rm MD}$ increases, which in turn improves $E_C$. 
For instance, $N^*_{\rm gNB}$ goes from around $16.2$ to around $2.4$ for $P_T=18$~dBm as $P_{\rm MD}$ increases from $0$ to $0.1$.
This is because a non-zero misdetection probability allows for more UEs to violate the SNR constraint, so the gNB can reduce the number of antennas and the transmission power to save energy, and still make the problem feasible.
In addition, as $N^*_{\rm gNB}$ decreases, the corresponding beamwidth increases, so the range of feasible values of $v$ and $T_{\rm SS}$ to maintain beam alignment increases, 
which also increases the size of the feasibility regions. %This is also evident from the fact that 
For instance, $P_T\geq 15$~dBm is feasible for $P_{\rm MD}=0$, while $P_T\geq 10$~dBm is feasible for $P_{\rm MD}=0.1$.
%As the probability of misdetection increases, the constraint \eqref{C4} in the optimization problem is relaxed. 
Nevertheless, $N^*_{\rm gNB}$ and $P_t^*$ eventually saturate when increasing $P_{\rm MD}$. In particular, $N^*_{\rm gNB}$ tends to the minimum feasible value ($N^*_{\rm gNB}\approx2$), whereas the asymptotic value of $P_t^*$ decreases as $P_{\rm MD}$ increases. 
%Accordingly, %it can be observed that 
%the value of $P_T$ corresponding to the $E_C$ minima decreases with $P_{\rm MD}$, whereas it increases with the number of users $K$.

\begin{figure}[t!]
	\centering 
    \setlength\fwidth{0.75\columnwidth}
    \setlength\fheight{0.3\columnwidth}
	\subfigure[$K=200$.] {% This file was created by matlab2tikz.
%
%The latest updates can be retrieved from
%  http://www.mathworks.com/matlabcentral/fileexchange/22022-matlab2tikz-matlab2tikz
%where you can also make suggestions and rate matlab2tikz.
%
\definecolor{mycolor1}{rgb}{0.00000,0.44700,0.74100}%
\definecolor{mycolor2}{rgb}{0.85000,0.32500,0.09800}%
\definecolor{mycolor3}{rgb}{0.92900,0.69400,0.12500}%
\definecolor{mycolor4}{rgb}{0.46667,0.67451,0.18824}%
\begin{tikzpicture}

\begin{axis}[%
width=\fwidth,
height=\fheight,
at={(0\fwidth,0\fheight)},
scale only axis,
bar shift auto,
xmin=0.509090909090909,
xmax=4.49090909090909,
xtick={1,2,3,4},
xticklabels={{8},{16},{32},{64}},
xlabel={$N_{\rm SS}$},
ymin=0,
ymax=160,
ytick={5,20,40,80,160},
ylabel={$T_{\rm SS}$ (ms)},
axis background/.style={fill=white},
xmajorgrids,
ymajorgrids,
legend style={at={(0.02,0.62)}, legend columns=2, anchor=south west, legend cell align=left, align=left, draw=white!15!black}
]
\addplot[ybar, bar width=0.145, fill=mycolor1, draw=black, area legend] table[row sep=crcr] {%
1	10\\
2	20\\
3	40\\
4	80\\
};
\addlegendentry{$P_{\rm MD}=0$}

\addplot[ybar, bar width=0.145, preaction={fill, mycolor2}, pattern={north east lines}, draw=black, area legend] table[row sep=crcr] {%
1	20\\
2	20\\
3	80\\
4	160\\
};
\addlegendentry{$P_{\rm MD}=0.01$}

\addplot[ybar, bar width=0.145, preaction={fill, mycolor3}, pattern={dots}, draw=black, area legend] table[row sep=crcr] {%
1	20\\
2	40\\
3	80\\
4	160\\
};
\addlegendentry{$P_{\rm MD}=0.05$}

\addplot[ybar, bar width=0.145, preaction={fill, mycolor4}, pattern={grid}, draw=black, area legend] table[row sep=crcr] {%
1	20\\
2	40\\
3	80\\
4	160\\
};
\addlegendentry{$P_{\rm MD}=0.1$}
\end{axis}

\end{tikzpicture}%
\label{FR_K200}} 
	\subfigure[$K=500$.] {% This file was created by matlab2tikz.
%
%The latest updates can be retrieved from
%  http://www.mathworks.com/matlabcentral/fileexchange/22022-matlab2tikz-matlab2tikz
%where you can also make suggestions and rate matlab2tikz.
%
\definecolor{mycolor1}{rgb}{0.00000,0.44700,0.74100}%
\definecolor{mycolor2}{rgb}{0.85000,0.32500,0.09800}%
\definecolor{mycolor3}{rgb}{0.92900,0.69400,0.12500}%
\definecolor{mycolor4}{rgb}{0.46667,0.67451,0.18824}%
\begin{tikzpicture}

\begin{axis}[%
width=\fwidth,
height=\fheight,
at={(0\fwidth,0\fheight)},
scale only axis,
bar shift auto,
xmin=0.509090909090909,
xmax=4.49090909090909,
xtick={1,2,3,4},
xticklabels={{8},{16},{32},{64}},
xlabel={$N_{\rm SS}$},
ymin=0,
ymax=160,
%ymode=log,
ytick={5,20,40,80,160},
ylabel={$T_{\rm SS}$ (ms)},
axis background/.style={fill=white},
xmajorgrids,
ymajorgrids,
legend style={at={(0.02,0.62)}, legend columns=2, anchor=south west, legend cell align=left, align=left, draw=white!15!black}
]
\addplot[ybar, bar width=0.145, fill=mycolor1, draw=black, area legend] table[row sep=crcr] {%
1	5\\
2	10\\
3	20\\
4	40\\
};
\addlegendentry{$P_{\rm MD}=0$}

\addplot[ybar, bar width=0.145, preaction={fill, mycolor2}, pattern={north east lines}, draw=black, area legend] table[row sep=crcr] {%
1	5\\
2	10\\
3	20\\
4	80\\
};
\addlegendentry{$P_{\rm MD}=0.01$}

\addplot[ybar, bar width=0.145, preaction={fill, mycolor3}, pattern={dots}, draw=black, area legend] table[row sep=crcr] {%
1	10\\
2	20\\
3	40\\
4	80\\
};
\addlegendentry{$P_{\rm MD}=0.05$}

\addplot[ybar, bar width=0.145, preaction={fill, mycolor4}, pattern={grid}, draw=black, area legend] table[row sep=crcr] {%
1	10\\
2	20\\
3	40\\
4	80\\
};
\addlegendentry{$P_{\rm MD}=0.1$}
\end{axis}

\end{tikzpicture}%
\label{FR_K500}}
	\caption{Feasibility regions for $P_T=18$ dBm, $\tau=7$ dB, $v=5$ m/s, and $K\in\{200, 500\}$, considering $P_{\rm MD}\in\{0, 0.01, 0.05, 0.1\}$.}\label{FR_pmd}
\end{figure}
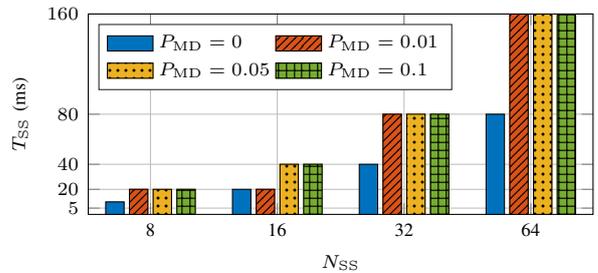
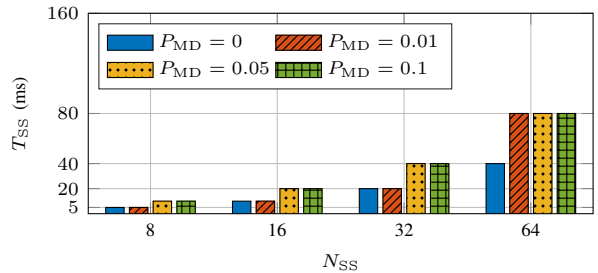

Tables~\ref{FR_pmd_table_K200} and~\ref{FR_pmd_table_K500} %corroborate these findings, 
report the upper bound for the feasible product $v T_{\rm SS}$, for $P_{\rm MD}$ ranging from $0$ to $0.1$ and $K\in\{200, 500\}$. 
According to the results in Sec.~\ref{sec_FL}, 
the feasibility region becomes smaller as $K$ increases for $P_{\rm MD}=0$. %as observed in Fig.~\ref{FR_K}. 
Conversely, the impact of $K$ %does not impact 
on the feasibility region becomes negligible when $P_{\rm MD}>0$. 
%which prevents the misalignment of the UEs with respect to the corresponding associated beam also increases.
On one side, as $K$ increases there would be more UEs at the cell edge, with an SNR which is close to that of the worst case. On the other side, $P_{\rm MD}>0$ would likely mitigate the SNR constraint especially on those worst-case UEs, so that the corresponding feasibility regions would remain unchanged if the two effects are balanced. 
Fig.~\ref{FR_pmd} %and~\ref{FR_K500} 
depicts the %plot of the 
feasibility regions for $P_T=18$~dBm, $\tau=7$~dB,  %at UE mobility of 
$v=5$ m/s, and $K\in\{200, 500\}$, and corroborates these results. 
%The feasibility regions have been depicted as bar plots here in order to distinguish between the overlapping regions. 
We observe that for $P_{\rm MD}=0$ and $N_{\rm SS}=8$, 
the upper bound of the burst periodicity goes from $T_{\rm SS}=10$ to $5$~ms as $K$ increases, while for $P_{\rm MD}>0$ the limits remain unchanged. 
A similar behavior can be observed as a function of $N_{\rm SS}$. %as well. %The expanding feasibility limit with increasing $N_{\rm SS}$ values is also apparent.

%\begin{Definitions}{}{}
\Summary
As $P_{\rm MD}>0$, the system is less constrained in terms of SNR, which improves the energy consumption.
Also, the size of the feasibility regions tends to increase, so the set of feasible values of $v$ and $T_{\rm SS}$ also increases.
Finally, the impact of $K$ is negligible.
%which prevents the misalignment of the UEs with respect to the corresponding associated beam also increases.
%\end{Definitions}

\begin{table}[t!]
	\centering
	\renewcommand{\arraystretch}{1.1}
	\caption{Feasibility regions for $P_T=18$~dBm, $\tau=7$ dB, and $K=200$.}
	\begin{tabular}{ c|c|c|c|c }
		\hline\
		& \multicolumn{4}{c}{Upper bound on $vT_{\rm SS}$} \\
		\hline
		\diagbox{$N_{\rm SS}$}{$P_{\rm MD}$} & $0$ &$0.01$ & $0.05$ & $0.1$ \\
		\hline
		$8$   &$0.08$ m  &$0.32$ m  & $2.24$ m & $4.8$ m\\
		\hline
		$16$   &$0.16$ m &$0.64$ m  & $4.8$ m & $4.8$ m \\
		\hline
		$32$   &$0.32$ m & $1.6$ m & $4.8$ m & $4.8$ m \\
		\hline
		$64$   &$0.64$ m & $4.8$ m & $4.8$ m & $4.8$ m \\
		\hline
	\end{tabular}
	\label{FR_pmd_table_K200}
 %\vspace{-0.33cm}
\end{table}

\begin{table}[t!]
	\centering
	\renewcommand{\arraystretch}{1.1}
	\caption{Feasibility regions for $P_T=18$~dBm, $\tau=7$ dB, and $K=500$.}
	\begin{tabular}{ c|c|c|c|c }
		\hline\
		& \multicolumn{4}{c}{Upper bound on $vT_{\rm SS}$} \\
		\hline
		\diagbox{$N_{\rm SS}$}{$P_{\rm MD}$} & $0$ &$0.01$ & $0.05$ & $0.1$ \\
		\hline
		$8$   &$0.03$ m  &$0.24$ m  & $2.24$ m & $4.8$ m\\
		\hline
		$16$   &$0.06$ m &$0.56$ m  & $4.8$ m & $4.8$ m \\
		\hline
		$32$   &$0.14$ m & $1.28$ m & $4.8$ m & $4.8$ m \\
		\hline
		$64$   &$0.32$ m & $4.8$ m & $4.8$ m & $4.8$ m \\
		\hline
	\end{tabular}
	\label{FR_pmd_table_K500}
 %\vspace{-0.33cm}
\end{table}

\subsection{Design Guidelines}

%In Fig.~\ref{ECT_{avg}} we plot $\overline{E_C}_{t}$ as a function of $T_{\rm SS}$ and $N_{\rm SS}$. We can see that the energy consumption increases when $T_{\rm SS}$  ($N_{\rm SS}$) decreases (increases), as the corresponding number of beam management operations per unit of time increases. 
The analysis in Sec.~\ref{sec_FL} suggests that %, generally speaking, 
decreasing the duration of beam management,
%as stated in Eq.~\eqref{OP2},
i.e., reducing $T_{\rm SS}$ and increasing $N_{\rm SS}$, promotes feasibility.
%would be the optimal configuration (i.e., in the bottom-right part of the feasibility regions) to make the problem feasible. 
%This corresponds to the bottom-right portion of the feasibility region depicted in Fig.~\ref{FR_K}. 
%This would imply. 
However, this approach also leads to an increase of the beam management overhead, i.e., more time-frequency resources are used for sending the control signals, and fewer resources are available for data transmission~\cite{giordani2019tutorial}. 
In fact, even though $E_C$ is independent of the values of $v$, $T_{\rm SS}$ and $N_{\rm SS}$ (see Eq.~\eqref{EC1}), and is thus constant within the feasibility region, $P_C$ is not, as per Eq.~\eqref{ECT}. Therefore, for a given feasible solution $\{N^*_{\rm gNB}$, $P_t^*\}$, we can choose $T_{\rm SS}$ and $N_{\rm SS}$ so as to optimize the trade-off between beam management overhead and the resulting energy consumption.
In particular, we propose to operate in the top-left part of the feasibility regions, that is choosing the highest feasible $T_{\rm SS}$ and the lowest feasible $N_{\rm SS}$, %, i.e., maximum feasible burst periodicity at $N_{\rm SS}=8$ 
which therefore represents the optimal beam management configuration for RedCap.
This choice still satisfies Constraint $C_1$ ($C_1^{'}$) of P1 (P2) in terms of SNR since we are within the feasibility regions.
Table~\ref{tradeoff_table} reports the optimal values of $T_{\rm SS}$ and $N_{\rm SS}$ for $P_T=18$~dBm, $\tau=7$~dB, $v=5$~m/s, $K\in\{200, 500\}$, and $P_{\rm MD}\in\{0, 0.01. 0.05. 0.1\}$.
%as obtained from Fig.~\ref{FR_pmd}.

%Please note that the maximum feasible burst periodicity decreases as the SNR threshold increases. An analogous effect would result from a drop in the transmit power $P_T$.

\begin{table}[t]
	\centering
	\renewcommand{\arraystretch}{1.1}
	\caption{Optimal RedCap beam management parameters %after trade-off at 
 for\\$P_T=18$~dBm and $\tau=7$~dB.}
	\begin{tabular}{ c|c|c|c|c }
		\hline
		& \multicolumn{4}{c}{Optimal ($N_{\rm SS}, T_{\rm SS}$)} \\
		\hline
		\diagbox{$K$}{$P_{\rm MD}$} & $0$ &$0.01$ & $0.05$ & $0.1$ \\
		\hline
		$200$   &$8,10$ ms  & $8,40$ ms  & $8,160$ ms & $8,160$ ms\\
		\hline
		$500$   &$8,5$ ms & $8,40$ ms & $8,160$ ms & $8,160$ ms \\
		\hline
	\end{tabular}
	\label{tradeoff_table}
 %\vspace{-0.33cm}
\end{table}

%\begin{Definitions}{}{}
\Summary
The optimal and feasible beam management configuration for RedCap in terms of both energy efficiency and overhead is to choose the highest possible $T_{\rm SS}$ and the lowest possible $N_{\rm SS}$ within the feasibility regions, which is turn determines the corresponding optimal values of $N^*_{\rm gNB}$ and $P_t^*$.
%which prevents the misalignment of the UEs with respect to the corresponding associated beam also increases.
%\end{Definitions}

\section{Conclusions and Future Work}\label{conclude}
\label{sec:concs}
The objective of this work is to minimize the energy consumption of 5G RedCap beam management in an \gls{sma} scenario, where both the base station and the end users are battery powered, and thus feature strict energy constraints.
%To this end, we derive the impact of the beam management parameters, the user speed, and the gNB configuration on the energy consumption.
To this end, we formalize an optimization problem to choose the number of antenna elements ($N^*_{\rm gNB}$) and the transmission power ($P_t^*$) at the gNB to minimize energy consumption while guaranteeing a sufficient link quality.
Since the problem is an MINLP, we develop a custom algorithm to estimate a feasible solution. 
%Specifically, we identify the number of antenna elements at the gNB as the dominating variable in the optimization problem. 
%Then, we develop an algorithm which, based on this rationale, estimates the solution of the full problem.

Numerical results show that the optimal number of antennas at the gNB decreases with the maximum transmission power, and increases when considering more demanding SNR constraints. 
Conversely, the optimal transmission power at the gNB depends on the number of UEs, and is proportional to both the maximum transmission power and the target minimum link quality. 
Notably, the solution of the optimization problem is independent of the users velocity ($v$) and the beam management parameters, i.e., the SSB burst periodicity ($T_{\rm SS}$) and the number of SSBs per burst ($N_{\rm SS}$).
%Overall, the total energy consumed ($E_C$) for sending the beam management control signals (SSBs) shows a non-monotonic trend with respec to the  $P_T$ and is minimum for minimum $\tau$. Therefore, we conclude that the $E_C$ at the gNB for sending the control signals is minimal at the smallest possible SNR threshold and at some intermediate transmission power. 
In addition, we identify via simulations, and bound analytically, the feasibility regions of the problem, i.e., the set of parameters for which the optimization problem is feasible.
%This feasibility limit shrinks with an increase in $K$ and expands with an increase in $P_{\rm MD}$. 
Finally, we observe a trade-off between the energy consumption and the beam management overhead, and propose the optimal feasible beam management configurations for RedCap to optimize the two effects.

As part of our future work, we plan to jointly optimize the energy consumption of both beam management and data transmission, as well as to consider more advanced optimization methods, for example based on machine learning.

\balance
\bibliographystyle{IEEEtran}
\bibliography{bibl}

% Generated by IEEEtran.bst, version: 1.14 (2015/08/26)
\begin{thebibliography}{10}
\providecommand{\url}[1]{#1}
\csname url@samestyle\endcsname
\providecommand{\newblock}{\relax}
\providecommand{\bibinfo}[2]{#2}
\providecommand{\BIBentrySTDinterwordspacing}{\spaceskip=0pt\relax}
\providecommand{\BIBentryALTinterwordstretchfactor}{4}
\providecommand{\BIBentryALTinterwordspacing}{\spaceskip=\fontdimen2\font plus
\BIBentryALTinterwordstretchfactor\fontdimen3\font minus \fontdimen4\font\relax}
\providecommand{\BIBforeignlanguage}[2]{{%
\expandafter\ifx\csname l@#1\endcsname\relax
\typeout{** WARNING: IEEEtran.bst: No hyphenation pattern has been}%
\typeout{** loaded for the language `#1'. Using the pattern for}%
\typeout{** the default language instead.}%
\else
\language=\csname l@#1\endcsname
\fi
#2}}
\providecommand{\BIBdecl}{\relax}
\BIBdecl

\bibitem{10437644}
M.~Rawat, M.~Pagin, M.~Giordani, L.-A. Dufrene, Q.~Lampin, and M.~Zorzi, ``Minimizing energy consumption for {5G NR} beam management for {RedCap} devices,'' in \emph{IEEE GLOBECOM, Kuala Lumpur, Malaysia}, 2023.

\bibitem{38300}
3GPP, ``{NR} and {NG-RAN} overall description,'' TS 38.300, 2018.

\bibitem{itu-r-2083}
ITU-R, ``{IMT Vision - Framework and overall objectives of the future development of IMT for 2020 and beyond},'' Recommendation ITU-R M.2083, Sep. 2015.

\bibitem{22261}
3GPP, ``{Service requirements for next generation new services and markets},'' TS 22.261, 2018.

\bibitem{atzori2010internet}
L.~Atzori, A.~Iera, and G.~Morabito, ``{The Internet of Things: A survey},'' \emph{Comput. Networks}, vol.~54, no.~15, pp. 2787--2805, May 2010.

\bibitem{zanella2014internet}
A.~Zanella, N.~Bui, A.~Castellani, L.~Vangelista, and M.~Zorzi, ``{Internet of Things for Smart Cities},'' \emph{IEEE Internet Things J.}, vol.~1, no.~1, pp. 22--32, Feb. 2014.

\bibitem{magrin2017performance}
D.~Magrin, M.~Centenaro, and L.~Vangelista, ``{Performance evaluation of LoRa networks in a smart city scenario},'' in \emph{IEEE ICC, Paris, France}, 2017.

\bibitem{matz2020systematic}
A.~P. Matz, J.-A. Fernandez-Prieto, J.~Ca{\~n}ada-Bago, and U.~Birkel, ``{A systematic analysis of narrowband IoT quality of service},'' \emph{Sensors}, vol.~20, no.~6, p. 1636, Mar. 2020.

\bibitem{ribeiro2018outdoor}
G.~G. Ribeiro, L.~F. de~Lima, L.~Oliveira, J.~J. Rodrigues, C.~N. Marins, and G.~A. Marcondes, ``{An outdoor localization system based on SigFox},'' in \emph{IEEE VTC Spring, Porto, Portugal}, 2018.

\bibitem{ayoub2018internet}
W.~Ayoub, A.~E. Samhat, F.~Nouvel, M.~Mroue, and J.-C. Pr{\'e}votet, ``{Internet of Mobile Things: Overview of LoRaWAN, DASH7, and NB-IoT in LPWANs standards and supported mobility},'' \emph{IEEE Commun. Surv. Tutorials}, vol.~21, no.~2, pp. 1561--1581, Secondquarter 2018.

\bibitem{8066090}
S.~S.~H. Hajjaj and K.~S.~M. Sahari, ``{Review of agriculture robotics: Practicality and feasibility},'' in \emph{IEEE IRIS, Tokyo, Japan}, 2016.

\bibitem{22804}
3GPP, ``{Study on Communication for Automation in Vertical domains (CAV) -- Release 15},'' TR 22.804, 2020.

\bibitem{38875}
------, ``{Study on support of reduced capability NR devices - Rel. 17},'' TR 38.875, 2020.

\bibitem{varsier20215g}
N.~Varsier, L.-A. Dufr{\`e}ne, M.~Dumay, Q.~Lampin, and J.~Schwoerer, ``A {5G} {New Radio} for balanced and mixed {IoT} use cases: {Challenges} and key enablers in {FR1} band,'' \emph{IEEE Commun. Mag.}, vol.~59, no.~4, pp. 82--87, May 2021.

\bibitem{pagin2023nrlight}
M.~Pagin, T.~Zugno, M.~Giordani, L.-A. Dufrene, Q.~Lampin, and M.~Zorzi, ``{5G NR-Light} at millimeter waves: {Design} guidelines for mid-market {IoT} use cases,'' \emph{IEEE ICNC, Honolulu, HI, USA}, 2023.

\bibitem{chiu2019active}
S.-E. Chiu, N.~Ronquillo, and T.~Javidi, ``Active learning and {CSI} acquisition for {mmWave} initial alignment,'' \emph{IEEE J. Sel. Areas Commun.}, vol.~37, no.~11, pp. 2474--2489, Nov. 2019.

\bibitem{9605254}
M.~Hussain and N.~Michelusi, ``Learning and adaptation for millimeter-wave beam tracking and training: {A} dual timescale variational framework,'' \emph{IEEE J. Sel. Areas Commun.}, vol.~40, no.~1, pp. 37--53, Jan. 2022.

\bibitem{scalabrin2018beam}
M.~Scalabrin, N.~Michelusi, and M.~Rossi, ``{Beam training and data transmission optimization in millimeter-wave vehicular networks},'' in \emph{IEEE GLOBECOM, Abu Dhabi, United Arab Emirates}, 2018.

\bibitem{8445969}
Y.~Wang, M.~Narasimha, and R.~W. Heath, ``{MmWave} beam prediction with situational awareness: {A} machine learning approach,'' in \emph{IEEE SPAWC Wrkshp, Kalamata, Greece}, 2018.

\bibitem{giordani2019tutorial}
M.~Giordani, M.~Polese, A.~Roy, D.~Castor, and M.~Zorzi, ``{A Tutorial on Beam Management for 3GPP NR at mmWave Frequencies},'' \emph{IEEE Commun. Surv. Tutorials}, vol.~21, no.~1, pp. 173--196, Firstquarter, 2019.

\bibitem{giordani2017improved}
M.~Giordani and M.~Zorzi, ``{Improved user tracking in 5G millimeter wave mobile networks via refinement operations},'' in \emph{IEEE Med-Hoc-Net, Budva, Montenegro}, 2017.

\bibitem{8590817}
X.~Zhao, A.~M.~A. Abdo, Y.~Zhang, S.~Geng, and J.~Zhang, ``Single {RF}-chain beam training for {MU-MIMO} energy efficiency and information-centric {IoT} millimeter wave communications,'' \emph{IEEE Access}, vol.~7, pp. 6597--6610, Dec. 2018.

\bibitem{9773088}
N.~Zeulin, A.~Ponomarenko-Timofeev, O.~Galinina, and S.~Andreev, ``{ML}-assisted beam selection via digital twins for time-sensitive industrial {IoT},'' \emph{IEEE Internet Things Mag.}, vol.~5, no.~1, pp. 36--40, Mar. 2022.

\bibitem{mukherjee2020energy}
A.~Mukherjee, ``Energy-efficient beam management in millimeter-wave shared spectrum,'' \emph{IEEE Wireless Commun.}, vol.~27, no.~5, pp. 38--43, Oct. 2020.

\bibitem{ericcson2023}
Ericsson, ``{RedCap} - expanding the {5G} device ecosystem for consumers and industries,'' Ericcson White Paper, 2023.

\bibitem{traspadini2023energy}
A.~Traspadini, M.~Giordani, G.~Giambene, T.~De~Cola, and M.~Zorzi, ``{On the Energy Consumption of UAV Edge Computing in Non-Terrestrial Networks},'' in \emph{Asilomar Conference on Signals, Systems, and Computers, Pacific Grove, CA, USA}, 2023.

\bibitem{3GPPrel16}
3GPP, ``{Study on channel model for frequencies from 0.5 to 100 GHz- Release 16},'' \emph{TR 38.901}, 2020.

\bibitem{rawat2023optimal}
M.~Rawat, M.~Giordani, B.~Lall, A.~Chaoub, and M.~Zorzi, ``{On the Optimal Beamwidth of UAV-Assisted Networks Operating at Millimeter Waves},'' \emph{IEEE WCNC, Glasgow, United Kingdom}, 2023.

\bibitem{abbas2017}
W.~B. Abbas, F.~Gomez-Cuba, and M.~Zorzi, ``Millimeter wave receiver efficiency: {A} comprehensive comparison of beamforming schemes with low resolution {ADCs},'' \emph{IEEE Trans. Wireless Commun.}, vol.~16, no.~12, pp. 8131--8146, Dec. 2017.

\bibitem{balanis}
C.~A. Balanis, \emph{Antenna theory: analysis and design}, 4th~ed.\hskip 1em plus 0.5em minus 0.4em\relax John Wiley \& Sons, Inc, 2015.

\bibitem{energyefficiency2018}
L.~N. Ribeiro, S.~Schwarz, M.~Rupp, and A.~L.~F. de~Almeida, ``Energy efficiency of {mmWave Massive MIMO} precoding with low-resolution {DACs},'' \emph{IEEE J. Sel. Top. Signal Process.}, vol.~12, no.~2, pp. 298--312, Apr. 2018.

\bibitem{poweramp2017}
K.~Greene, A.~Sarkar, and B.~Floyd, ``A 60-{GHz} dual-vector {Doherty} beamformer,'' \emph{IEEE J. Solid-State Circuits}, vol.~52, no.~5, pp. 1373--1387, May 2017.

\bibitem{giordani2018coverage}
M.~Giordani, M.~Rebato, A.~Zanella, and M.~Zorzi, ``{Coverage and connectivity analysis of millimeter wave vehicular networks},'' \emph{Ad Hoc Networks}, vol.~80, pp. 158--171, Nov. 2018.

\bibitem{misener2014antigone}
R.~Misener and C.~A. Floudas, ``{ANTIGONE: Algorithms for coNTinuous/Integer Global Optimization of Nonlinear Equations},'' \emph{J. Global Optim.}, vol.~59, no. 2-3, pp. 503--526, Mar. 2014.

\bibitem{gurobi}
\BIBentryALTinterwordspacing
{Gurobi Optimization, LLC}, ``{Gurobi Optimizer Reference Manual},'' 2023. [Online]. Available: \url{https://www.gurobi.com}
\BIBentrySTDinterwordspacing

\end{thebibliography}

\end{document}